\documentclass[manuscript]{aastex}
\usepackage{graphicx}
\usepackage{epstopdf}

\shorttitle{Models for 15 Galactic Supernova Remnants}
\shortauthors{Leahy \& Ranasinghe}

\begin{document}


\title{Evolutionary Models for 43 Galactic Supernova Remnants with Distances and X-ray Spectra}


\author{D.A. Leahy, S. Ranasinghe. M. Gelowitz}
\affil{Department of Physics $\&$ Astronomy, University of Calgary, Calgary,
Alberta T2N 1N4, Canada}



\begin{abstract}
{The X-ray emission from a supernova remnant (SNR) is a powerful diagnostic of the state of the shocked plasma.
The temperature (kT) and the emission measure (EM) of the shocked-gas are related to the energy of the explosion, the
age of the SNR, and the density of the surrounding medium. 
Progress in X-ray observations of SNRs has resulted in a significant sample of Galactic SNRs with measured  kT and EM values.
We apply spherically symmetric SNR evolution models to a new set of 43 SNRs to estimate  ages, explosion energies, and circum-stellar medium densities.
The distribution of ages yields a SNR birth rate. 
The energies and densities are well fit with log-normal distributions, with wide dispersions.
SNRs with two emission components are used to distinguish between SNR models with uniform ISM and with
stellar wind environment. We find type Ia SNRs to be consistent with a stellar wind environment. 
Inclusion of stellar wind SNR models has a significant effect on estimated lifetimes and explosion energies of
SNRs. This reduces the discrepancy between the estimated SNR birthrate and the SN rate of the Galaxy.}
\end{abstract}


\keywords{supernova remnants:}

\section{Introduction}

Supernova remnants (SNRs) have a great impact in astrophysics, including injection of energy (e.g. \citealt{2005Cox}) 
and newly synthesized elements into the interstellar medium (e.g. \citealt{2012Vink}). 
Valuable constraints for stellar evolution and the evolution of the interstellar medium and the Galaxy can
be obtained from SNR studies. 
The goals of SNR research include understanding explosions of supernovae (SN) and the resulting energy 
and mass injection into the interstellar medium.

Approximately 300 SNRs have been observed in our Galaxy by their radio emission \citep{2019Green}.
Only a small number have been physically characterized, including determination of evolutionary state, 
explosion energy and age. 
Several historical SNR have modelled with hydrodynamic simulations (e.g. \citealt{2006Badenes}).
Most SNRs have less complete observations than the historical SNRs, thus lack a definite age, and have 
not been modelled in detail.
The non-historical SNRs are the main part of the Galactic SNR population. 
For many of these, a simple Sedov model with an assumed energy and ISM density has been applied.
However, better modelling is required to derive energies, densities and ages from observations.
Thus to study the physical properties of the SNR population it is necessary to use models simpler than 
full hydrodynamic modelling, but which include more physical effects than the Sedov model.

For the purpose of characterization of SNRs, we have developed a set of 
SNR models for spherically symmetric SNRs, which are based on hydrodynamic calculations. The most recent
version of the models are described in \citet{2019AJ....158..149L}. 
The model calculates temperatures and emission measures of the hot plasma, for forward-shocked and for 
reverse-shocked gas, based on hydrodynamic simulations.
The resulting SNR models facilitate the process of using observations to estimate SNR physical properties.

This work includes the following. 
We extend the solution to the inverse problem, which calculates the initial
parameters of a SNR from its observed properties, for the latest set of models in \citet{2019AJ....158..149L}.
We apply this  inverse solution to a set of 43 SNRs with distances and with observed temperatures and emission measures.
The structure of the paper is as follows.
In Section~\ref{sec:model} we present an overview of the SNR models and the solution of the inverse problem. 
In Section~\ref{sec:sample}  the current sample of 43 SNRs is described. 
The results of model fits are given in Section~\ref{sec:SNRmodels} for 3 cases: with a standard model for all SNRs;
with a cloudy SNR model for mixed morphology SNRs; and with a general model for two-component SNRs .
In Section~\ref{sec:properties} we discuss the results, including statistical properties of the Galactic SNR population.
The conclusions are summarized in Section~\ref{sec:conclusion}.

\section{SNR Evolution and Our Adopted Model}\label{sec:model}

A SNR is the interaction of the SN  ejecta with the interstellar medium (ISM). 
The various stages of evolution of a SNR are labelled the 
ejecta-dominated stage (ED), the adiabatic or Sedov-Taylor stage
(ST), the radiative pressure-driven snowplow (PDS) and the radiative momentum-conserving shell (MCS).
These stages are reviewed in, e.g., \cite{1988cioffi}, \cite{1999truelove} (hereafter TM99), 
\citet{2012Vink} and \citet{2017LeahyWilliams}.
In addition, there are the transitions between stages, called ED to ST, ST to PDS, and PDS to MCS, respectively.
The ED to ST stage is important because the SNR is still bright, and it is long-lived enough that a significant fraction of SNRs are
likely in this phase.

For simplicity our models assume that the SN ejecta and ISM are spherically symmetric.
The ISM density profile is a power-law centred on the SN explosion, given by $\rho_{ISM}=\rho_s r^{-s}$, 
with s=0 (constant density medium) or s=2 (stellar wind density profile).
The unshocked ejectum has a power-law density  $\rho_{ej}\propto r^{-n}$.
With these profiles, the ED phase of the SNR evolution has a
self-similar evolution (\citealt{1982Chev} and \citealt{1985Nad}).
The evolution of  SNR shock radius was extended for the ED to ST and ST phases by TM99.

The model for SNR evolution that we use is partly based on the TM99 analytic solutions, with additional
features. 
A detailed description of the model is given in \citet{2019AJ....158..149L} and \citet{2017LeahyWilliams}. 
We incorporate Coulomb collisional electron heating, consistent with the observational results of \cite{2013Ghav}. 
The emission measure ($EM$) and emission measure-weighted temperature ($T_{EM}$) are calculated for ED, ED to ST 
and ST phases. $EM$ and $T_{EM}$ are calculated from the interior structure of the SNR from hydrodynamic simulations.
 $dEM_{FS}$, $dT_{FS}$, $dEM_{RS}$ and $dT_{RS}$ are calculated separately for forward shocked (FS) gas and for reverse shocked (RS) gas.
 Analytic fits in terms of piecewise powerlaw functions are made \citep{2019AJ....158..149L}
  to the numerically determined $dEM_{FS}$, $dT_{FS}$, $dEM_{RS}$ and $dT_{RS}$.  
The inverse problem is solved, which takes as input the SNR observed properties and determines the 
initial properties of the SNR.

\begin{figure*}[ht!]
\includegraphics[width=\textwidth]{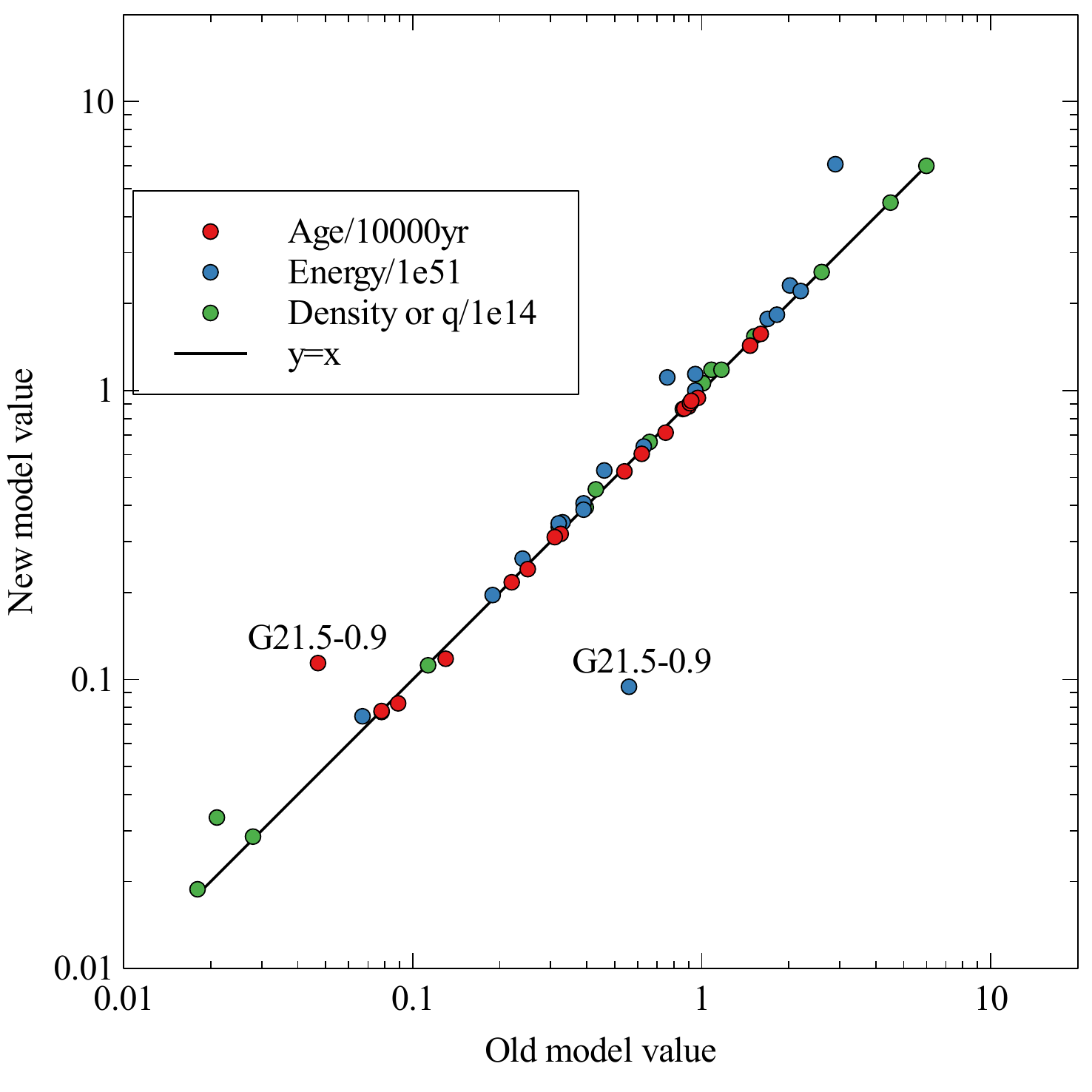}%
\caption{Comparison of fits using our new SNR model \citep{2019AJ....158..149L}   to fits using the older model for the 15 SNRs analyzed by \citet{2018ApJ...866....9L}.
Model inputs are radius, emission measure and temperature from Table 1 of
 \citet{2018ApJ...866....9L}. Outputs are age, energy and density in both cases (see text for details). 
Table 4 of \citet{2018ApJ...866....9L} gives other inputs for the model: model type (forward shock or reverse shock), external density power-law (s=0 for ISM, s=2 for stellar wind),  progenitor envelope power-law (n, in the range 6 to 14), and ejected mass. The ``y=x'' line indicates exact agreement between the two models. 
The outlier SNR G21.5-0.9 is discussed in the text.
}
\label{fig:compare}
\end{figure*}

To illustrate the changes in the current model  \citep{2019AJ....158..149L} compared to the model used by 
\citet{2018ApJ...866....9L}, we recalculate the inverse models for the 15 SNRs in \citet{2018ApJ...866....9L}.
The main difference in the models is that our newer model includes accurate calculation of $dEM_{RS}$ and $dT_{RS}$
for the stages after the ED stage, rather than a simple linear interpolation.
The parameters from the two models applied to the 15 SNRs are compared in Figure~\ref{fig:compare}.
Perfect agreement between the two models is indicated by the black line.
The derived ages, energies and densities from the new model are very close to those from the previous simpler model.   
One clear exception is for the young SNR G21.5-0.9: there is a significant change in derived parameters because 
that SNR is in the ED to ST transition stage at a time where the difference between the newer accurate treatment 
and the previous linear interpolation is the largest.

\subsection{The inverse problem: application of models to SNR data}

The forward model, using initial SNR conditions and calculating conditions at time t, is described in \citet{2019AJ....158..149L}.
The input parameters of the SNR forward model are: age;
SN energy $E_0$; ejected mass $M_{ej}$; ejecta power-law index n; ejecta composition;
ISM temperature $T_{ISM}$; ISM composition; ISM power-law index s (0 or 2); and
ISM density $n_0$ (if s=0) or mass-loss parameter $\rho_s=\frac{\dot{M}}{4\pi v_w}$ (if s=2).
For Galactic SNRs, we take the ISM to be solar composition. 
We take the composition of ejecta and the mass of ejecta  to be fixed, 
with values depending on the type of SN. For  Type Ia we use $1.4M_{\odot}$  ejecta mass 
and for core-collapse (CC) or unknown type we use $5M_{\odot}$. 
We use the CC (for CC and unknown types) and Ia ejecta abundances from \citet{2018ApJ...866....9L}. 
$T_{ISM}$ ,which doesn't affect the evolution unless the SNR is very old, is set to 100 K.
The resulting number of free parameters is 5 for the forward model.

The inverse model is solved by an iterative procedure. The observed FS radius ($R_{FS}$), FS
emission measure ($EM_{FS}$) and  FS EM weighted temperature ($T_{FS}$) for a given SNR
are used to estimate ISM density, explosion energy and SNR age. The forward model is applied to calculate
$R_{FS,model}$, $EM_{FS,model}$ and $T_{FS,model}$. 
Then the differences between the input and model value are used to determine new density, energy and age
values. The process is repeated until the density, energy and age reproduce the input 
$R_{FS}$, $EM_{FS}$ and $T_{FS}$ to better than 1 part in $10^4$. To determine errors in the 
model parameters, the inverse solution is repeated for several sets of input parameters, for
which the observed values are replaced in turn by their upper and lower limits.

To handle the fact that the inverse model has 3 input parameters, we apply the
inverse solution for each $s$, $n$ case separately. For each fixed $s$, $n$ case there are 3 free parameters
of the forward model (see first paragraph of this section). Thus, we can obtain a unique solution
of SNR density, energy and age for each $s$, $n$. 

Because the smooth dependence of the SNR models on $n$, we chose to limit the number of $n$ cases to
$n=7$ and $n=10$.
In cases where the observational data to distinguish these cases do not exist, we choose the $s=0$,  $n=7$ model.
In cases where observations on reverse-shocked ejecta exists,
we choose the $s$, $n$ case which most closely reproduces the reverse shock properties ($EM_{RS}$ and $T_{RS}$).
 
\section{The SNR Sample}\label{sec:sample}

The basic parameters of the SNRs were obtained from the catalogue of Galactic SNRs  \citep{{2019Green}} and 
the catalogue of High Energy Observations of Galactic Supernova Remnants 
(\url{http://www.physics.umanitoba.ca/snr/SNRcat/} \citep{Ferrand 2012}). 
The SNRs were chosen to have reliable distances and X-ray observations. 
The average of major and minor axes (from \cite{{2019Green}}) was used to estimate the average outer shock radius as input to the spherical SNR model. 

Out of the 294 SNRs \citep{{2019Green}}, \cite{2018ApJ...866....9L} analysed 15 SNRs covered by the Very Large Array (VLA) Galactic
Plane Survey (VGPS). 
The sample in the current work is comprised of 43 SNRs that lie outside the region that is covered by the VGPS. 
For each SNR, we obtained the temperature of the shocked gas from and calculated the emission measure from the
reference for the X-ray spectrum and its spectral fit. 
For many cases, only a part of the SNR had an X-ray spectrum, so we had to extrapolate the
given EM to account for the rest of the SNR. 
The EM values were adjusted from the published values, which assumed a distance, 
 to the best-measured distance in the literature, and the dependence on distance is specified.
 In the next section are notes relevant to the modelling for each of the 43 SNRs, in order of increasing Galactic
 longitude. 
 The adopted values of $EM$ and $kT$ are given in Table~\ref{tab:TBLobserved}. 

We carry out three sets of models on the SNR sample. 
For the first set of models,
we fit the data with a single hot plasma component, which is taken to be emission from gas heated
by the forward shock (FS) and set $s=0$, $n=7$. 
The second set of models is applied for the subset of seven probable mixed morphology SNRs.
For these SNRs a more appropriate model is that for a SNR in a cloudy ISM (\citealt{1991WL}, hereafter WL). 
We have implemented this in our SNR modelling code, with the additional calculations of including
electron-ion temperature equilibration, and of solving the inverse problem for the cloudy model.
The third set of models is carried out for those 12 SNRs with two measured hot plasma components.
The two components are likely from the forward shocked gas and the reverse shocked gas.
The reverse shocked gas is identified by its enriched element abundances, if measured.
For all 12 cases, the emission measure of the forward shocked, $EM_{FS}$,  gas exceeds 
that of the reverse shocked gas.
Thus, we choose $R_{FS}$, $EM_{FS}$ and $kT_{FS}$ as model inputs and compute predicted
$EM_{RS}$ and $kT_{RS}$ for different $s$, $n$ cases. 

 
\subsection{Notes on  Individual SNRs in the Sample}\label{subsec:indSNR} 

\textbf{G38.7-1.4:} This is a mixed morphology SNR and is large in radius, $\simeq$15 pc.
The Chandra ACIS-I X-ray spectrum of G38.7-1.4 is well described by an absorbed 
non collisional-ionization-equilibrium (NIE) plasma model \citep{2014Huang} 
with no evidence for RS heated ejecta.
Based on the ACIS-I image, the spectral extraction area used by \citet{2014Huang} includes $\simeq$0.5
of the X-ray counts from the SNR, thus we multiply their EM by 2 to obtain an estimate of the total 
EM of the SNR. 

\textbf{G53.6-2.2:} This  is a large (radius $\simeq$35 pc), mixed morphology SNR.
 \citet{2015Broersen} analyzed the ACIS-I X-ray spectrum of the whole SNR and found a two-component spectrum:
 one from FS-shocked ISM and one from RS-shocked ejecta

\textbf{G67.7+1.8:} This  is a small (radius$\simeq$4 pc), mixed morphology SNR which likely hosts a central compact 
object (CCO).
\citet{2009Hui} extract ACIS-I spectra for the Northern Rim and the Southern Rim, and fit them with a collisional ionization equilibrium (CIE) model, finding significant enhancements in Mg, Si and S (at
$\sim$ 3, 3 and 15 times solar abundance).
Based on the ACIS-I image of the SNR and the given spectral extraction regions, we estimate $\simeq$0.33
of the X-ray flux of the SNR was included in the spectral extraction regions, so multiply their fit EM by a 
factor of 3 to obtain the total EM for the SNR.

\textbf{G78.2+2.1:} This was observed with ACIS-I and ACIS-S, with spectral and image analysis by \citet{2013Leahy}.
The NEI spectrum model APEC as fit to spectra of 5 different regions, which yields the mean temperature.
The whole area of the SNR was not covered by the ACIS observations, so we use the norm of the 
APEC fit to the ROSAT whole SNR spectrum to obtain the EM. 

\textbf{G82.2+5.3/ W63:} The X-ray spectrum was obtained with the ASCA GIS detector and analyzed with 
the VMEKAL model by \citet{2004Mavromatakis}. The GIS data only covers a small central region of
the SNR, so we use the ROSAT PSPC spectral fits of the southern area of the SNR to estimate 
a scaling factor of 10 to obtain the EM for the whole SNR.

\textbf{G84.2-0.8:} The X-ray spectrum was obtained with ACIS-I and analyzed by \cite{2012Leahy},
using an APEC model. We compared the area observed in X-ray by ACIS-I with the radio image
to estimate a scaling factor of 2.5 to obtain the EM for the whole SNR from the ACIS fit value.

\textbf{G85.4+0.7 and G85.9-0.6:} These two SNRs were observed using XMM Newton PN and MOS detectors 
and the spectra were analyzed by \cite{2008Jackson} using a VPSHOCK model. The spectrum
extraction region encompasses the whole SNR in both cases so no adjustment to EM was necessary.
Both G85.4+0.7 and G85.9-0.6 are mixed morphology SNRs. 

\textbf{G89.0+4.7/ HB21:} The X-ray spectrum was obtained with ASCA SIS and GIS instruments 
by \citet{2006Lazendic} and analyzed with the VNEI model. From the ROSAT PSPC image of 
G89.0+4.7, the flux outside of the ASCA spectrum extraction regions were estimated to
yield a scaling factor of 2.0 to obtain the EM for the whole SNR.

\textbf{G109.1-1.0/ CTB109:} The spectrum was obtained with the Suzaku XIS instrument and analyzed
by \citet{2015Nakano}. The emission of the NE and SE regions, which avoids the bright central 
object 1E2259+586, was analyzed using a two component NEI plus VNEI model. 
The NEI model represented shocked ISM and was solar abundance; the VNEI model represented 
``shocked ejecta'' and showed very mild enhancements of Si and S (by factors of 1.0 to 1.7). 
By examining the Suzaku XIS mosaic image of G109.1-1.0 and comparing the X-ray extraction regions 
to the radio image, we estimate a factor of  2.0 correction to obtain the EM for the whole SNR.

\textbf{G116.9+0.2/ CTB 1:} The ASCA SIS and GIS spectra of this mixed morphology SNR 
were analyzed by \citet{2006Lazendic}. The largest areas analyzed were called W-whole and NE-whole,
and were fit with a two-component VRAY (CIE) model. The ROSAT PSPC image shows that the
sum of X-ray emission of these two overlapping regions is exceeds the total X-ray emission of the
SNR by about 50\%, so we multiply the sum of the two EMs by 0.66 to estimate the total SNR EM.

\textbf{G132.7+1.3/ HB3:} This was observed in X-rays with ROSAT/PSPC, ASCA GIS and SIS and XMM-Newton pn
detectors, and analyzed by \citet{2006Lazendic} using a one-component VRAY model. Of the 5 overlapping
regions, the 3 central regions show enhanced Mg abundance, by factor of $\simeq$2, and the
central XMM-pn region shows additional O enhancement by factor of $\simeq$2.
We use the average of temperature of all 5 regions and the ROSAT PSPC image to estimate 
the correction factor of 1.5 for EM, accounting for regions not covered by the spectra and accounting 
for the overlap of GIS, SIS and XMM-pn in the central region.

\textbf{G156.2+5.7:} Observations by the Suzaku XIS instrument were analyzed by \cite{2009Katsuda}.
The ROSAT PSPC image show the locations of the XIS fields on this X-ray shell-type SNR. 
Spectral parameters were derived using a VNEI model for the regions E1, E2, E3, E4 and NW4.
The  NW1, NW2 and NW3  regions were fit using two component VNEI +NEI or VNEI + 
power-law models. The central ellipse region was best fit using three component 
2VNEI +NEI or 2VNEI + power-law models, with one of the VNEI components, with enhanced
abundances of Si and S, identified as ejecta emission. We adopt the ejecta VNEI component from
the central ellipse as the ejecta kT and EM. For the ISM component,  we use the average temperature
of the other regions. 
\cite{2009Katsuda} quote EM as the line-of-sight integral $EM_{los}=\int n_e n_H dl$, 
related to surface brightness, instead of the usual volume integral $EM=\int n_e n_H dV$, 
related to total SNR emission. 
 The regions observed with Suzaku were of 
average surface brightness for this SNR, as determined by comparison with the ROSAT/PSPC image.  
Thus we average their $EM_{los}$ values (omitting the ejecta component),
then convert to a total ISM EM using the area of X-ray emission for the whole SNR from the ROSAT/PSPC image.

\textbf{G160.9+2.6/ HB9:} This was observed by the ROSAT/PSPC and analyzed by \citet{1995Leahy}.
Table 2 gives best fit single component Raymond-Smith model spectral parameters for 7 regions covering the
SNR. The average temperature and sum of the EMs was used for the total SNR emission.

\textbf{G166.0+4.3/ VRO 42.05.01:} This mixed morphology SNR was observed with Suzaku XIS instrument.
X-ray spectra covering a large part of the NE region and a large part of the W region of the SNR were
analyzed by \citet{2017Matsumura}. Five different models were applied to the NE spectrum and one to the
W spectrum. All models were the recombining plasma emission model VVRNEI, but differed in which
abundances were assumed fixed at solar. We adopt the best fit models w-i and ne-iii, and use the
average temperature and summed EM, with a correction by a factor of 1.4 for the emission from the
SNR not covered by the X-ray extraction regions. 

\textbf{G260.4-3.4/ Puppis A/ MSH 08-44:} This was observed with XMM-Newton MOS.
Three regions (North, West and South) were extracted for spectral analysis by \citet{2013Katsuda}.
Their case A VNEI fits assumed Fe/H to be solar and gave better fits than their case B VNEI fits, 
which assumed O/H of 2000 times solar. 
The case A fits give enhanced O, Ne, Mg and Si by factors of $\simeq$3 to 10.
We use the average T of the 3 regions. We sum the EMs, and use a correction factor of 2.0 to
account for SNR emission outside the extraction regions, which was 
 obtained from the Chandra ACIS-I image of the SNR.

\textbf{G272.2-3.2:} This is classified as a non-thermal composite SNR, and was observed with ASCA and ROSAT.  
The whole SNR spectrum from ASCA GIS and ROSAT PSPC was analyzed by \citet{2001Harrus} using a NEI model with no enhanced abundance. We adopt their T and EM to represent the shocked ISM for this SNR. 
Small regions A and B from the ASCA SIS were fit by the NEI model and gave similar T, and lower EM
consistent with the smaller region size.

\textbf{G296.7-0.9:} This was observed by XMM-Newton MOS and analyzed by \citet{2013Prinz}.
The spectrum of the X-ray bright SE rim was fit by various models, with the NEI, PSHOCK, SEDOV and
GNEI models yielding good fits. We adopt the T and EM parameters from the NEI model, and correct
the EM by a factor of 3.0 to account for the SNR emission outside the extraction region.   

\textbf{G296.8-0.3:} The whole SNR was observed with XMM-Newton MOS and analyzed by \cite{2012SanchezAyaso}.
Point-like sources were identified, including one object that is consistent with being a CCO, then excluded
from the SNR elliptical extraction region. The resulting SNR spectrum
was fit with a PSHOCK model, and we adopt those parameters.

\textbf{G299.2-2.9:} This is a type Ia SNR. \citet{2014Post} analyzed the Chandra ACIS-I spectrum of this
SNR, using 4 small extraction regions labelled North(N), South(S), West(W) and Shell(Sh), and the VPSHOCK model. 
N, S and W are all interior to the rim, with N and S close to the center of the SNR.
Enhanced abundances of Si S and Fe by factors of  4-8, 5-20, and 3-6, respectively, were 
found for N, S and W indicating the emission is from shocked ejecta. 
The Sh emission is taken as from shocked ISM. 
We use the Chandra image to estimate the correction
factors to apply to both the ejecta and ISM components to get whole SNR values:
$EM_{ej}\simeq 20EM_{ej,S}+20EM_{ej,N}+25EM_{ej,W}$ for the ejecta,
and $EM_{ISM}\simeq20EM_{ISM,S}+20EM_{ISM,N}+25EM_{ISM,W}+40EM_{ISM,Sh}$ for shocked ISM.

\textbf{G304.6+0.1/ Kes 17} This was observed with 
ASCA GIS and SIS and analyzed by \citet{2014Pannuti}.
G304.6+0.1 was also observed with XMM Newton MOS1, MOS2 and PN instruments, and the
extraction region covered the whole SNR. 
Various spectral models were fit by \citet{2014Pannuti}, with the VAPEC plus power-law giving the best fit.
This model had enhanced abundance of Mg by factor $\sim$10, and of Si and S by factors of $\sim$5.
The filled center X-ray morphology indicates G304.6+0.1 is likely a mixed morphology SNR. 

\textbf{G306.3-0.9:} This was observed with XMM Newton MOS1, MOS2 and pn, and with Chandra ACIS-I and
ACIS-S \citep{2016combi}. 3 X-ray point sources were excluded from the SNR spectra. Five
extraction regions for the XMM-Newton data covered $\simeq$60\% of the area of the SNR.
MOS1, MOS2 and PN spectra were extracted for the five regions and fit with VAPEC plus VNEI models.
The VAPEC component had mildly subsolar abundances and represents the shocked ISM component.
The VNEI component for NE, NW, C and SW regions showed enhancements of Si, S Ar, Ca and Fe,
by factors of 9-20 for NE, 2-9 for NW, 10-46 for C, 6-34 for SW and 1-2.6 for S. This indicates that
all but region S are likely dominated by shocked ejecta emission.  We adopt the VNEI for shocked
ejecta component and VAPEC for shocked ISM component, and correct the EM by factor 2.0 
for the area of the
SNR outside the spectral extraction regions.

\textbf{G308.4-1.4:} This was classed as a mixed morphology SNR based on a low resolution ROSAT image. It was 
observed with Chandra ACIS-I at high resolution by \citet{2012Hui}, showing it rather to be
a limb-brightened partial shell SNR in X-rays. ACIS spectra were extracted for six regions: two Outer
Rim regions, two Inner Rim regions and two Central regions. The regions pairs were combined for
fitting and fit with an NEI model with similar T but much large EM for the Outer Rim, which is significantly
brighter in X-rays. There is no evidence for enhanced abundances, so the emission is taken to
be shocked ISM. We use the EM-weighted average of the temperatures. The EM was summed, 
with correction factor of 2 to account for the area and brightness outside the extraction regions,
from the X-ray brightness image from Chandra.

\textbf{G309.2-0.6:} This was observed by the ASCA GIS and SIS instruments and analyzed by \citet{2001Rakowski}.
The SNR is shell type in radio, and centrally filled in X-rays. The ASCA SIS spectrum was
modelled with a VNEI model, and showed enhanced Si abundance (factor $>$34) only, with other 
heavy elements (Ne through Fe) solar or subsolar. The abundances do not give a clear indicator
of the SN type.
The X-ray point source in G309.2-0.6 has absorption 
consistent with belonging to the foreground star cluster NGC5281, so is not likely a compact object
associated with the SNR. 

\textbf{G311.5-0.3}: The whole SNR was analyzed by \citet{2017Pannuti} using ASCA and XMM-Newton observations.
The X-ray emission is centrally concentrated showing the SNR to be in the mixed morphology class.
The ASCA GIS+SIS and XMM MOS1+MOS2+PN spectra were fit with an APEC model. 

\textbf{G315.4-2.3/ RCW86:} This  was observed with XMM-Newton with RGS and MOS instruments,
 and analyzed by \citet{2014Broersen}. Analysis of X-ray line ratios with RGS was used to 
 constrain T and ionization age $n_{e}t$ values for 4 different regions along the rim of the 
 X-ray shell-type SNR.
The RGS plus MOS spectra of 3 regions (SW, NW, SE) were fit with a two component
VNEI model, one of low T and the second of high T, while the region NE was fit with a single
low T VNEI component. The low T component showed mildly subsolar
abundances, whereas the high T component showed enhanced O, Ne, Si and Fe, with variations.
O and Ne were highest (18 and 3.5 times solar) for the SW region, Fe was highest for the NW
region (24 times solar). We use the average T and sum of EMs for the low T component to represent 
shocked ISM and the average T and sum of EMs for the high T component to represent shocked
ejecta. 
The EMs were further corrected, based on the X-ray image of the SNR, for the emission not
included in the four spectral extraction regions, as follows: 
$EM_{total}\simeq5EM_{NE}+2.5EM_{NW}+4EM_{SW}+6EM_{SE}$.

\textbf{G322.1+0.0:} This has been associated with the high mass X-ray binary Circinus X-1, thus is
a CC type SNR. Chandra ACIS-I observations of G322.1+0.0 were analyzed by \citet{2013Heinz}
showing a shell type SNR surrounding the bright central jet from Circinus X-1.
The spectrum was fit with a SEDOV model. 
They quoted the line-of-sight $EM_{los}=\int n_e n_H dl$ 
which we converted to the usual volume integral $EM=\int n_e n_H dV$, using the area of the SNR.

\textbf{G327.4+0.4/ Kes 27:}, was observed with the Chandra ACIS-I 
instrument and analyzed by \citet{2008Chen}. It was previously thought to be
a  thermal composite-type SNR, but the Chandra image (Fig. 4 of \citealt{2008Chen}) 
shows a number of point sources and concentrations of bright diffuse emission much 
more consistent with a shell morphology in X-rays. 10 regions were chosen for spectrum
extraction, along the rim and in the interior of the SNR to study spectral variations. 
The spectrum of the whole SNR was fit with XSPEC VPSHOCK and VSEDOV models. The latter
gave a better fit and we adopt the parameters of that fit.

\textbf{G330.0+15/ Lupus Loop:} The whole SNR was observed by the HEAO-1 A2 LED detectors and 
analyzed by \citet{1991ApJ...374..218L}. We adopt the T and EM from the 
two-temperature Raymond-Smith model which was fit to the spectrum. The distance is
highly uncertain, and could be as near as 500 pc or as far as SN1006, at 1.7 kpc  
\citep{1991ApJ...374..218L}.

\textbf{G330.2+1.0:} This has a CCO and is a CC-type SNR with a clear shell of X-ray emission. 
XMM-Newton MOS1 and MOS2 and Chandra ACIS-I observations were analyzed by \citet{2009Park}.
Three small regions (NE, E and SW) were selected for extraction of spectra. 
SW and NE regions were fit with a power-law spectrum and E region was fit with
pshock or a pshock plus power-law model. Thus, SW and NE regions are dominated by emission consistent
with synchrotron, whereas E is consistent with shocked gas or a mixture of synchrotron and 
shocked gas emission. We adopt the parameters from the pshock plus power-law model.
The ISM EM was corrected, based on the X-ray image of the SNR, for the emission not
included in the E spectral extraction region:
$EM\simeq10EM_{E}$.

\textbf{G332.4-0.4/ RCW 103:} This was observed with Chandra ACIS-I and ACIS-S and analyzed
by \citet{2015Frank}. G332.4-0.4 is of shell type in X-rays and has a CCO. 
27 regions were chosen for spectral analysis, and fit with a VPSHOCK model.
The regions were classified as being dominated by shocked CSM  (16 regions)
or by shocked ejecta (11 regions). The ejecta regions were identified by enhancements in
abundance of Mg, Si, S, or Fe. Typical abundances for the shocked ejecta regions had
Ne in the range of 0.7-1.4 times solar, Mg 0.5-3.5 times solar, Si 0.5-4.6 times solar, 
S 0.4-1.5 times solar and Fe 0.8 to 6.8 times solar.
The temperature of the shocked ISM and shocked ejecta were taken as the averages 
of T values for all regions for the shocked ISM component or shocked ejecta component, respectively.
\citet{2015Frank} quote the line-of-sight $EM_{los}=\int n_e n_H dl$.
Thus we obtain the $EM_{los,ej,av}$ of the shocked ejecta and $EM_{los,ISM,av}$ of the shocked 
ISM components by averaging the $EM_{los}$ values of ejecta or ISM for all regions. 
These are then converted to the usual volume integral $EM=\int n_e n_H dV$, using the area of the SNR.

\textbf{G332.4+0.1/ MSH 16-51/ Kes 32:} This was observed with Chandra
ACIS-I \citep{2004Vink}. The SNR is shell-type in X-rays with a bright NW rim. 
 Their second fit (Method 2) corrected for foreground diffuse Galactic
 emission, so we adopt the parameters of that fit.
The spectral extraction region covered $\sim$1/3 of the area of the SNR and included
 the bright region, 
 with only $\sim10$\% of the emission outside the extraction region. Thus, we apply a correction 
 factor of 1.1 to obtain EM of the whole SNR.

\textbf{G332.5-5.6:} The central part was observed with the Suzaku XIS instrument 
\citep{2015Zhu}. The radio image shows three stripes of radio emission from the SNR
shell. The central radio stripe is covered by the Suzaku XIS field of view and is coincident with
the X-ray emission, indicating the X-ray emission is probably from the SNR shell. 
The XIS0 and XIS3 merged spectrum was fit with a VNEI plus power-law model.
O, Mg and Fe abundances were fit, with the result that O and Fe were subsolar (at 0.6-0.8 times solar)
and Mg was enhanced (1.2 times solar). Thus, the X-ray emission is consistent
with shocked ISM. We apply a correction factor of 4 to EM to account for the area of the
SNR not covered by the spectrum extraction region.

\textbf{G337.2-0.7:} This was observed with XMM-Newton MOS1, MOS2, PN and Chandra ACIS-S 
instruments \citep{2006Rakowski}.  The whole SNR spectrum was obtained by fitting the
data from all four detectors and fitting with a VNEI plus power-law model. 
Enhanced abundances for the VNEI component were derived for Ne, Mg, Si, S, Ar and Ca, 
with factors ranging from 1.7 to 12.
This indicates that the emission is either from shocked ejecta or a combination of shocked ejecta
and shocked ISM. 
Eleven small subregions ranging in location from rim to center were extracted for the four 
detectors. Each region was fit jointly (all four detectors) with a VNEI model, confirming that enhanced
Si, S and Ca exists for all eleven subregions. The enhancements are more characteristic of a type Ia
spectrum but the absence of enhanced Fe is harder to explain with a type Ia model \citep{2006Rakowski}. 

\textbf{G337.8-0.1/ Kes 41:} This thermal composite was observed with XMM-Newton \& Chandra 
 \citep{2015Zhang}. An elliptical region covering the brightest X-ray emission near
the center of the SNR was used for spectrum extraction for XMM MOS+PN and Chandra ACIS
data. The spectra were fit jointly with a VNEI model  with enhanced S and Ar abundances of
1.8 and 2.4 times solar, respectively. The small enhancements indicate the shocked gas is dominated
by shocked ISM. We adopt their T and EM, with a correction factor of 2 to EM to account for emission outside
the spectrum extraction area. 

\textbf{G347.3-0.5/ RX J1713.7-3946:} This synchrotron-dominated SNR had its thermal spectrum detected
by \cite{2015Katsuda} using XMM Newton MOS1+MOS2 and Suzaku XIS observations.
A small region of $\simeq$4 arcmin radius near the center, but excluding the CCO, was used
for spectrum extraction. We adopt their simultaneous spectral fit to MOS and XIS data using 
Model B with background region 2. The fit used the VVRNEI model  and gave mildly enhanced 
abundances of Mg, Si and Fe, by factors of 2-4.3 times solar. Thus the emission is likely mainly from
 shocked ISM. 
They quote the line-of-sight $EM_{los}=\int n_e n_H dl$.
Thus we convert to the usual volume integral $EM=\int n_e n_H dV$, using the area of the SNR.

\textbf{G348.5+0.1/ CTB 37A:} This was observed with XMM-Newton MOS1+MOS2+PN  and Chandra 
ACIS-I detectors and analyzed by \citet{2014Pannuti}.
The MOS1+MOS2+PN spectrum was extracted for the whole of G348.5+0.1, 
excluding the small bright hard spectrum region CXOU J171419.8-383023.
We adopt their best fitmodel which used a VAPEC model.
The only element with non-solar abundance was Si with 0.48 times solar,  indicating
the emission is dominated by shocked ISM.
The high resolution ACIS-I  image shows the SNR to be of shell-type in X-rays. 
The ACIS-I spectra were extracted for CXOU J171419.8-383023, and the Northeast and Southeast
rim of SNR G348.5+0.1. CXOU J171419.8-383023 had a power-law spectrum
whereas Northeast and Southeast regions were fit with an APEC model that had T agree with
the XMM-Newton spectrum fit and EMs each smaller by a factor $\simeq$4 than the total SNR EM, 
consistent with the smaller extraction region sizes.  


\textbf{G348.7+0.3/ CTB 37B:} This was observed with Suzaku XIS and Chandra ACIS by \citet{2009Nakamura}.
The ACIS image shows the X-ray emission is separated into a point source plus diffuse emission.
The diffuse emission is separated into two areas, region 1 and region 2 for both XIS and ACIS observations.
Region 3 is seen in the XIS image but not in the ACIS image, so is likely a time variable active star.
Region 1 and 2 have significantly more counts in the XIS observation, so the XIS data were used for
spectral modelling, using a VNEI plus power-law spectra model. The T of region 1 and 2 are consistent
with each other, and we sum the EMs and apply a small correction factor of 1.1 to account for emission
outside the extraction regions.

\textbf{G349.7+0.2 and G350.1-0.3:} These two CC-type SNRs  were observed with Suzaku XIS by \cite{2014Yasumi}.
The extraction region for both of these small angular sized SNRs included the whole SNR.
The spectra of both remnants were fit with a solar abundance CIE component plus a VNEI component.
For G349.7+0.2, the Mg and Ni abundances were found to be 3.6 and 7 times solar, while the other
elements (Al, Si, S, Ar, Ca and Fe) were found to be either solar or sub-solar (0.6 to 1 times solar).
For G350.1-0.3, all the abundances (Mg, Al, Si, S, Ar, C,  Fe and Ni) were found to be above solar, 
with values ranging from 1.4 (Al and Fe) to 14 (for Ni). The CIE component is listed as the first component
(shocked ISM) in Table~\ref{tab:TBLobserved} and the VNEI component  is listed as the second
 component (shocked ejecta). However, the low abundances of the VNEI component for  G349.7+0.2
 indicate that the VNEI component my be a mixture of shocked ISM and ejecta.

\textbf{G352.7-0.1:} This was observed by XMM-Newton and Chandra and analyzed by
\citet{2014Pannuti}. With higher resolution than previous observations, the Chandra ACIS image show 
that this small angular diameter SNR is not centrally brightened but is more consistent with being
an X-ray shell type remnant than a mixed morphology SNR. The Eastern and Whole SNR spectra
were extracted from the XMM MOS1+MOS2+PN data and fit with a VNEI model. The Whole SNR
spectrum showed enhanced Si and S with abundances 2.3-3.5 times solar, thus is likely a mixture of
shocked ISM and shocked ejecta, and likely dominated by shocked ISM as pure shocked ejecta should have
much larger abundances. Spectra were extracted for 6 regions plus a Whole SNR region from the
 ACIS data and were fit by a one component VNEI model, by a two component NEI + VNEI model,
 and by a two component VNEI+ VNEI model.
 The ACIS Whole region one component model yielded a T and EM consistent with the XMM spectrum fit
 and also showed enhanced Si and S abundances.
The ACIS Whole region NEI+VNEI model was a significant improvement over the one component model.
The cool NEI component was solar abundance and represents shocked ISM, whereas the
hot VNEI component had Si and S enriched by factors of 3.4 and 9.4 over solar, so is likely
dominated by shocked ejecta.
The ACIS Whole region VNEI+VNEI model consisted of  a hot Si-rich component (7 times solar) and
a cool S-rich component (37 times solar), but the $\chi^2$ of the fit was not significanty better than
the NEI+VNEI  model. Thus, we adopt NEI+VNEI  fit values here.

\textbf{G355.6-0.0:} This was observed by Suzaku \citep{2013Minami}.
The Suzaku XIS image shows this SNR to be filled center, i.e. a mixed morphology SNR.
The spectrum extraction region was 6$^\prime$ by 4$^\prime$ (major axes), smaller than the
 8$^\prime$ by 6$^\prime$ SNR,
 but containing all the bright emission from the SNR. Thus we apply no correction factor to the EM.
 The background region was chosen to avoid the nearby bright
 X-ray emission from the star cluster NGC6383. We adopt the parameters from the spectrum fit 
 using background $a$, which used a VAPEC model.  This had Si, S, Ar and Ca enhanced by factors
 of 1.6, 3.1, 5.8 and 15 times solar, respectively. This may be a mixture of shocked ISM and shocked
 ejecta emission, but likely is dominated by shocked ejecta because of the high Ca abundance.

\textbf{G359.1-0.5:} The whole of this likely mixed morphology type was observed by Suzaku and analyzed by \citet{2011Ohnishi}. 
One component CIE and two component CIE models did not well fit the XIS spectrum. One or two
component underionized plasma models  also did not fit the XIS spectrum of G359.1-0.5. 
An overionized plasma spectrum model was a significantly better fit, so we adopt the parameters
from that model here. For that fit, Mg, Si and S were overabundant by factors of 3.4, 12, and 17,
indicating that the component is consistent with being dominated by shocked ejecta.

\section{Results of SNR Models}\label{sec:SNRmodels}

\subsection{The Standard Model}\label{subsec:standard}

For our standard case (Standard Model), we adopt $s=0$, $n=7$ for the ISM and ejecta density profiles.
For known or suspected Type Ia SNRs we set the ejecta mass $M_{ej}$=1.4$M_{\odot}$,
and for known or suspected CC SNRs we set $M_{ej}$=5$M_{\odot}$.
Because the rate of CC type SNRs is several times higher than that of Ia type, the unknown types are more 
likely to be CC type. Thus we set $M_{ej}$=5 $M_{\odot}$ for unknown types.
Further justification of the higher CC rate is found in the SNR types from Table~\ref{tab:TBLobserved}:
22 SNRs are type CC or CC? compared to 8 which are type Ia or Ia?. 
The results from applying this model are shown in Table~\ref{tab:TBLstdmodel}.
Derived ages, energies and densities and their uncertainties are given. 
Uncertainties were found by running models with the input parameters set to their upper and lower limits.

\begin{figure*}[ht!]
\includegraphics[width=\textwidth]{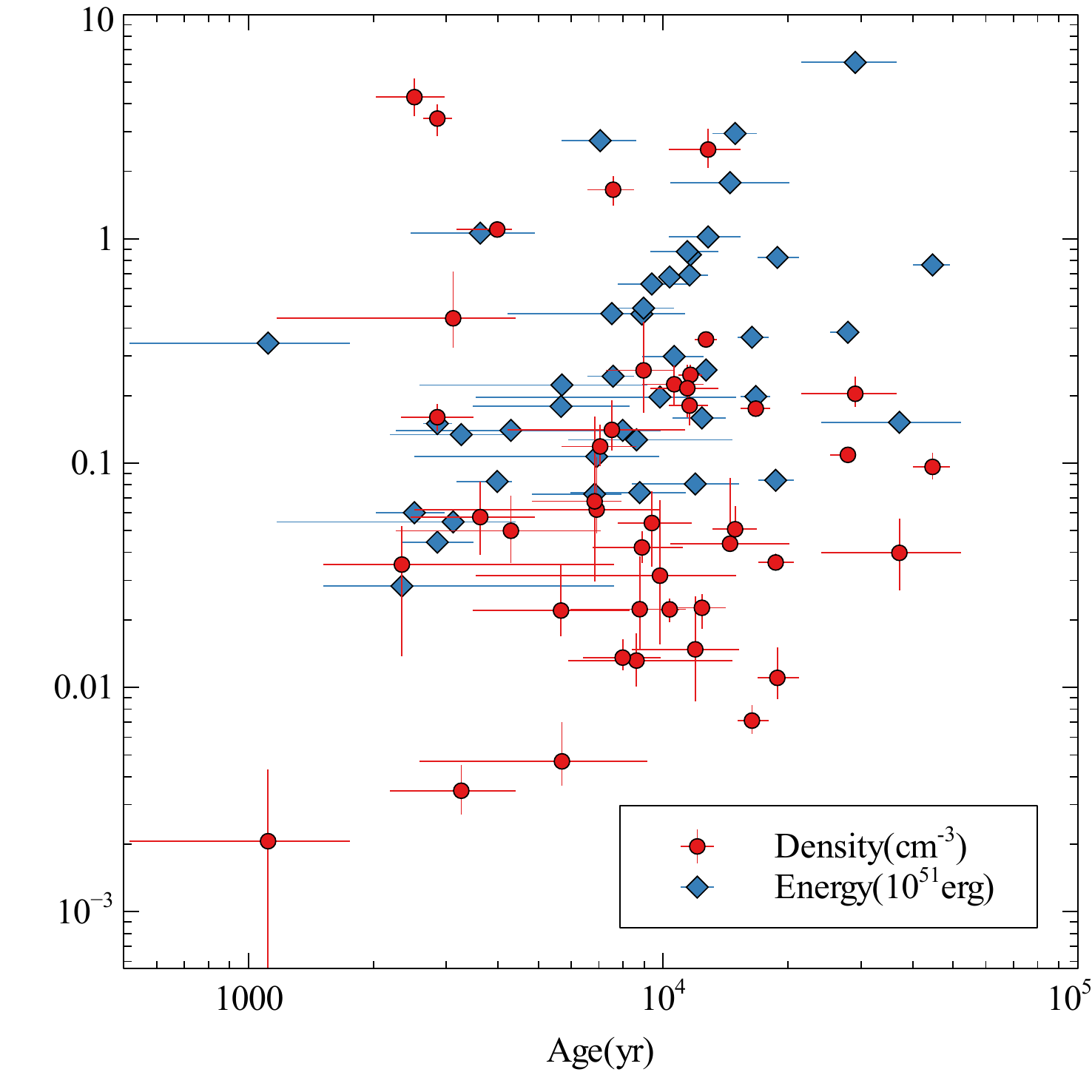}
\caption{Model energies and densities vs. model ages derived from the standard model (forward shock from an SNR with ISM and ejecta profiles with $s=0$, $n=7$).}
\label{fig:stdmodel}
\end{figure*}

The SNR ages range from 1100 yr to 36000 yr; the explosion energies range from 0.03 to 6 $\times10^{51}$ erg,
and the ISM densities range from 0.002 to 4.3 cm$^{-3}$.
Figure~\ref{fig:stdmodel} shows the explosion energy and ISM density plotted as functions of SNR age.
This illustrates the wide range of all three parameters and lack of correlation between them for the SNR sample.
The results of the standard model are discussed in Section~\ref{sec:snrstats} below.

\subsection{Mixed Morphology SNRs}\label{subsec:mmSNR}

Seven of the SNRs (G38.7-1.3, G53.6-2.2,  G82.2+5.3, G166.0+4.3, G304.6+0.1, G311.5-0.3 and G337.8-0.1) are mixed morphology SNRs.  
For these we run the cloudy (WL) model with $C/\tau=2$ (called the WL2 model).
The results are given in Table~\ref{tab:customWL} and discussed in Section~\ref{sec:cloudy} below.

\subsection{Two component SNRs}\label{subsec:2compSNR}

For 12 of the SNRs (see Table~\ref{tab:TBLobserved}), a second component is detected in the X-ray spectrum.
This  can be interpreted as reverse shock emission from the ejecta.
These SNRs are 
G53.6-2.2, G109.1-1.0, G116.9+0.2, G156.2+5.7, G299.2-2.9, G306.3-0.9,  
G315.4-2.3, G330.0+15.0, G332.4-0.4, G349.7+0.2, G350.1-0.3  and G352.7-0.1. 
For the CC types and unknown types, we assume $M_{ej}$=5$M_{\odot}$ and CC-type ejecta abundances. 
For the Ia types, we assume $M_{ej}$=1.4$M_{\odot}$ and Ia-type ejecta abundances. 

We apply our model to each of the 12 SNRs for 4 different cases of (s,n): (0,7), (0,10), (2,7) and (2,10).
The model returns the age, energy and density required in order to reproduce the forward shock
observed properties of $R_{FS}$, $EM_{FS}$ and $kT_{FS}$.
From the SNR model we calculate the predicted reverse shock properties $EM_{RS}$ and $kT_{RS}$.
Generally, the predicted EM-weighted reverse shock temperature decreases as n increases, and is larger
for s=0 than s=2.
In contrast, the predicted reverse shock emission measure increases as n increases, and is larger
for s=2 than s=0.
Table~\ref{tab:customS2} show the results of applying SNR models for different (s,n) cases.
The results from modelling the two-component SNRs are discussed in Section~\ref{sec:2comp} below.

\section{Discussion}\label{sec:properties}

\subsection{Standard Model}\label{sec:snrstats}


\subsubsection{Statistical properties}\label{sec:stats}

\begin{figure*}[ht!]
\plottwo{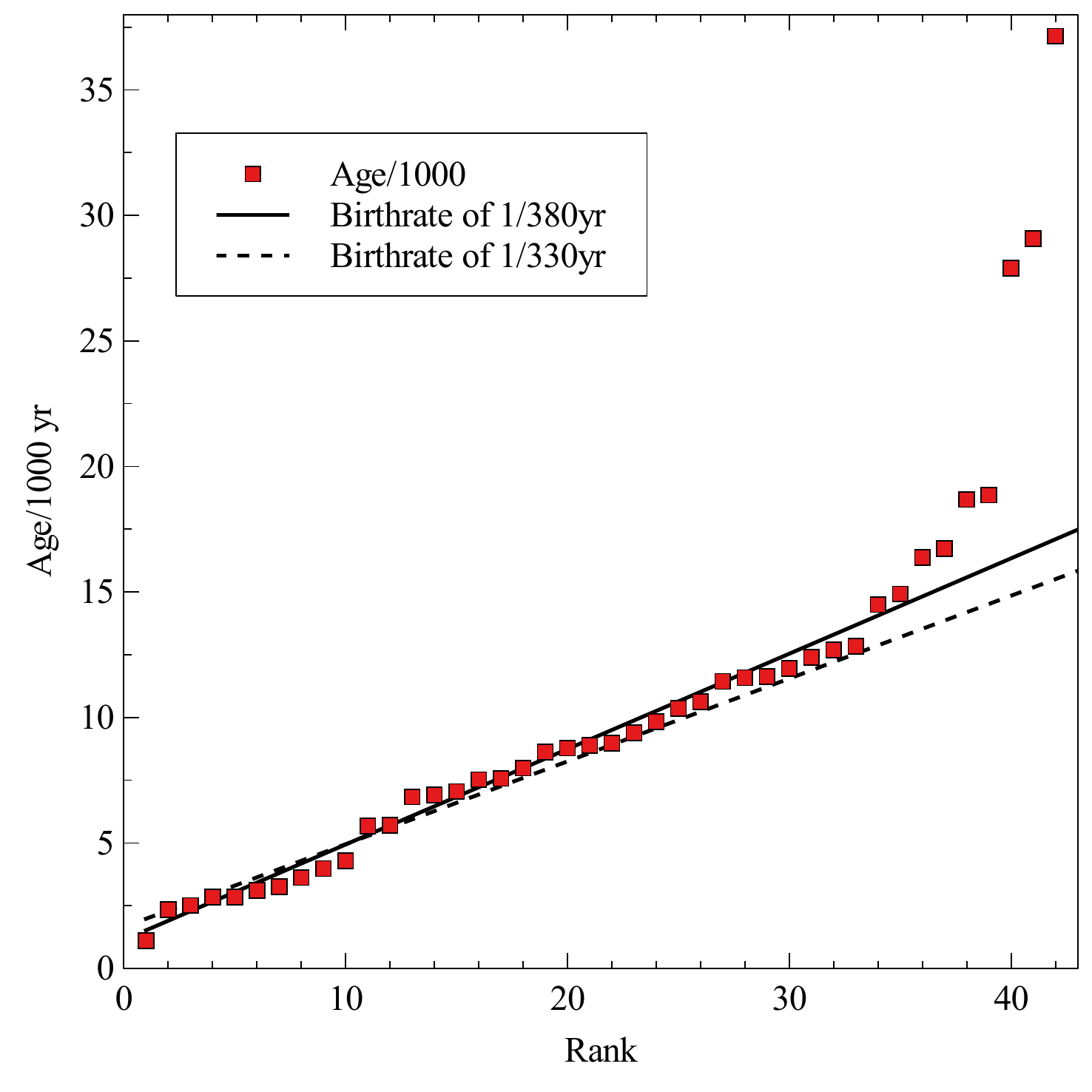}{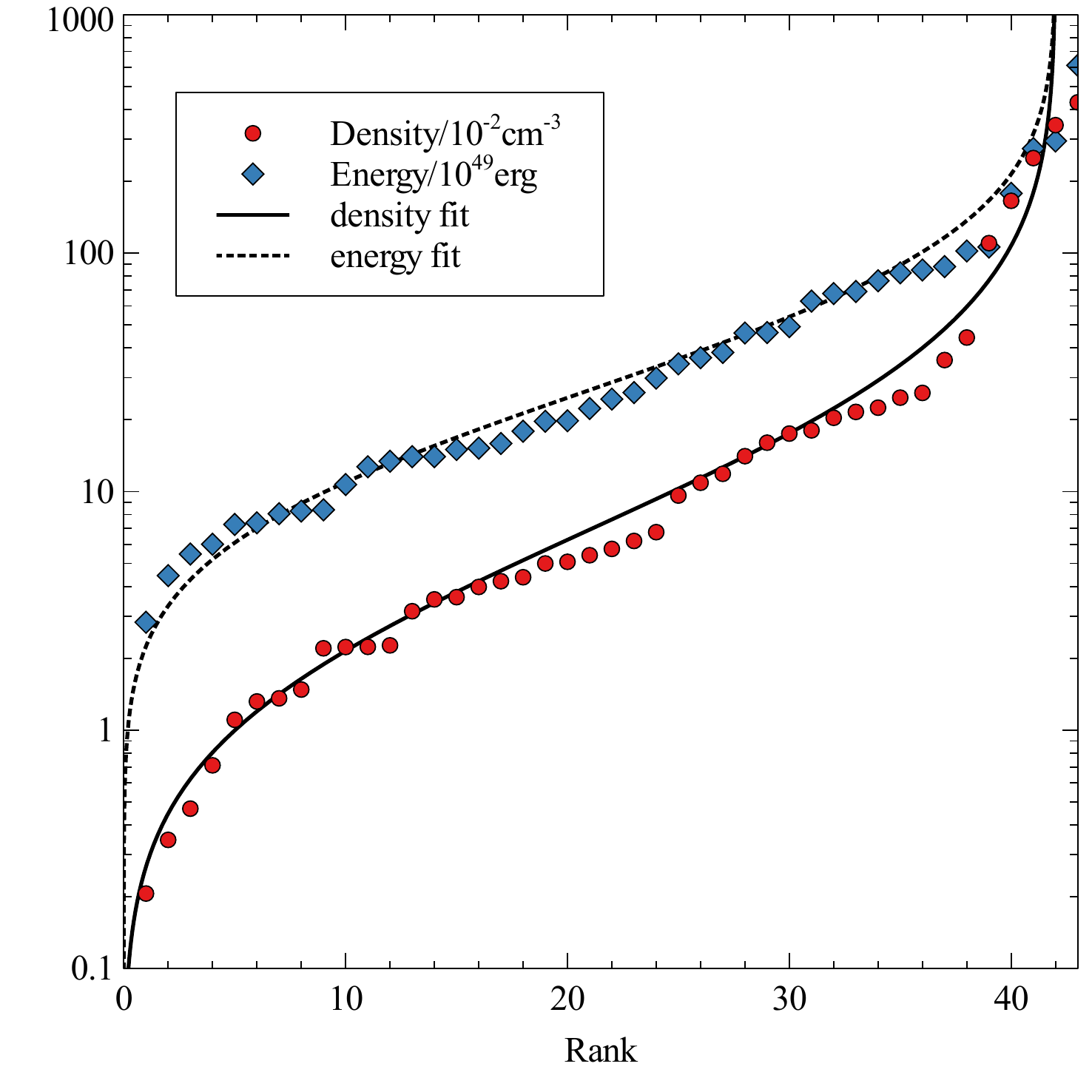}
\caption{Left panel:  Cumulative distribution of model ages from the standard model, and fit lines for birth rates of of 1 per 330 yr and 1 per 380yr. Right panel: Cumulative distributions of s=0, n=7 model  explosion energies and of densities
of the 43 SNRs. The fit lines are bset fit log-normal distributions. 
For explosion energies the mean is E$_0$=2.7$\times10^{50}$ erg and variance in log(E$_0$) is 0.54.
For densities the mean is $n_0$=6.9$\times10^{-2}$ cm$^{-3}$ and variance in log($n_0$) is 0.71.}
\label{fig:age_energy}
\end{figure*}

The statistical properties of the ages, explosion energies and ISM densities of the SNR sample are examined
using an analysis similar to that carried out for the LMC SNR population by \cite{2017Leahy} and
for 15 Galactic  SNRs \citep{2018ApJ...866....9L}. 
The left panel of  Figure~\ref{fig:age_energy} 
shows the cumulative distribution of model ages for our sample of 43 SNRs. 
The straight lines are the expected distributions for constant birth rates of 1/(330 yr) yr and  1/(380 yr), respectively. 
The fit lines have a non-zero x-intercepts of -5 and -3, respectively.
This is equivalent to allowing for 5 or 3 missing young age ($\lesssim$ 1000 yr) SNRs in the sample. 
We note that we did not include the set of  historical SNRs with age $\lesssim1000$ yr in our age distribution, so the non-zero x-intercept is expected.
Overall, a birthrate of 1/(350 yr) line is the best fit to the sample but values between 1/(380 yr) and 
1/(330 yr) are consistent with the age distribution.

The cumulative distribution of explosion energies is shown in  the right panel of  Figure~\ref{fig:age_energy}. 
This distribution is consistent with a log-normal distribution. 
The property that SNR explosion energies follow a log-normal distribution was discovered by \citet{2017Leahy}
and verified by  \citet{2018ApJ...866....9L}. 
Here, the best fit cumulative distribution has an average energy
$E_{av}= 2.7\times10^{50}$ erg and dispersion $\sigma_{logE}$=0.54 (a 1-$\sigma$ dispersion factor of 3.5).
These parameters are similar to those from the LMC SNR sample and from the 15 Galactic SNR sample. 

The cumulative distribution of ISM densities is shown in  the right panel of  Figure~\ref{fig:age_energy}. 
This distribution is consistent with a log-normal distribution, similar to that found by \citet{2017Leahy} and  \citet{2018ApJ...866....9L}.
The best fit cumulative distribution is shown by the solid line in  Figure~\ref{fig:age_energy}.  
The average density is $n_{0,av}= 0.069$ cm$^{-3}$ and dispersion is $\sigma_{log(n_0)}$=0.71 ( a 1-$\sigma$ dispersion factor 
of 5.1). 
For the LMC SNR sample, the mean density was similar and the dispersion was similar.
 The 15 Galactic SNRs subset   \citep{2018ApJ...866....9L} has a higher mean density and similar dispersion. 
 This is not surprising because the 15 Galactic SNRs were from the inner Galaxy where the density is expected to be higher.

\subsubsection{Uncertainties and systematic effects}\label{sec:systematics}

Some of the uncertainties of the analysis are mentioned here.
We have taken spectral fit results from the literature, where different SNRs have observations made
with different instruments, usually Chandra/ACIS, XMM/Newton MOS and PN, Suzaku XIS, 
or ASCA SIS and GIS instruments. There are effective area uncertainties of the
different instruments, which are at a level of $\sim$10\%. 
This can be compared to the errors in observed parameters of the SNRs. 
From Table~\ref{tab:TBLobserved}, distance and radius errors are typically 10 to 50\%, 
EM errors are typically 10 to 50\%, and kT errors typically 5 to 50\%.
Thus the observational errors dominate over instrument uncertainties.

In addition, many observations cover the whole SNR, but a significant number only cover 
a smaller region of the SNR.
The correction factors to EM that we applied were based on the total counts of  X-ray emission from
the SNR which are inside and outside the spectral extraction region. 
A crude estimate of the error in correction factor, $f$, is $\sqrt(f)$, which for many SNRs with correction factor of $\sim$2 results in an estimate of $\sim$40\% error. 
However, the counts in the X-ray image are better determined than this. 
E.g. typical SNR with 10,000 to 100,000 counts in the image has relative statistical error 
$\sqrt(10^{-4})=0.01$ to $\sqrt(10^{-4})=0.003$.
Thus the statistical error in estimating $f$ is very small and the error in $f$ is dominated by systematic errors.
Because the SNR measurement errors (see paragraph above) are large, they
probably dominate over errors in $f$.

\begin{figure*}[ht!]
\plottwo{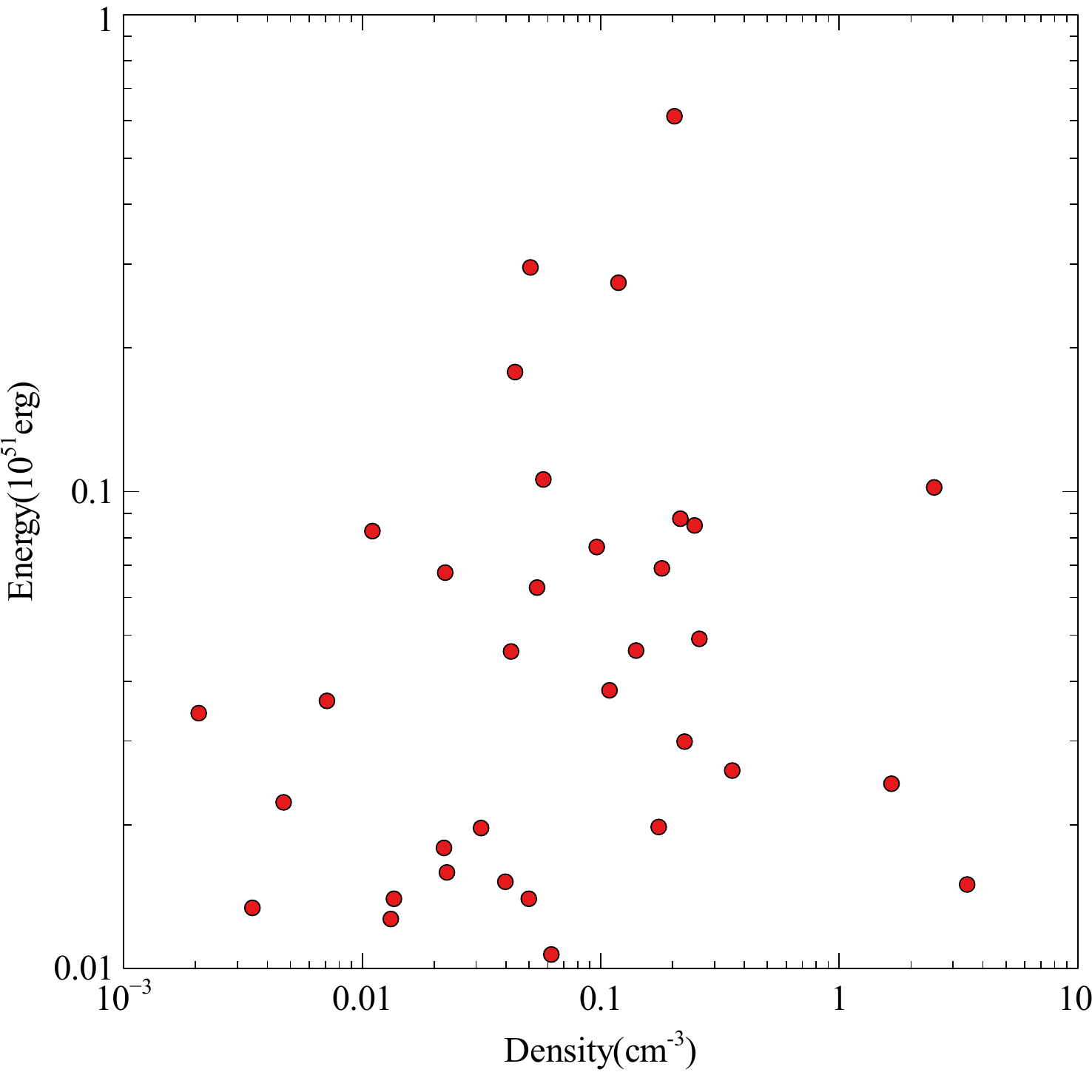}{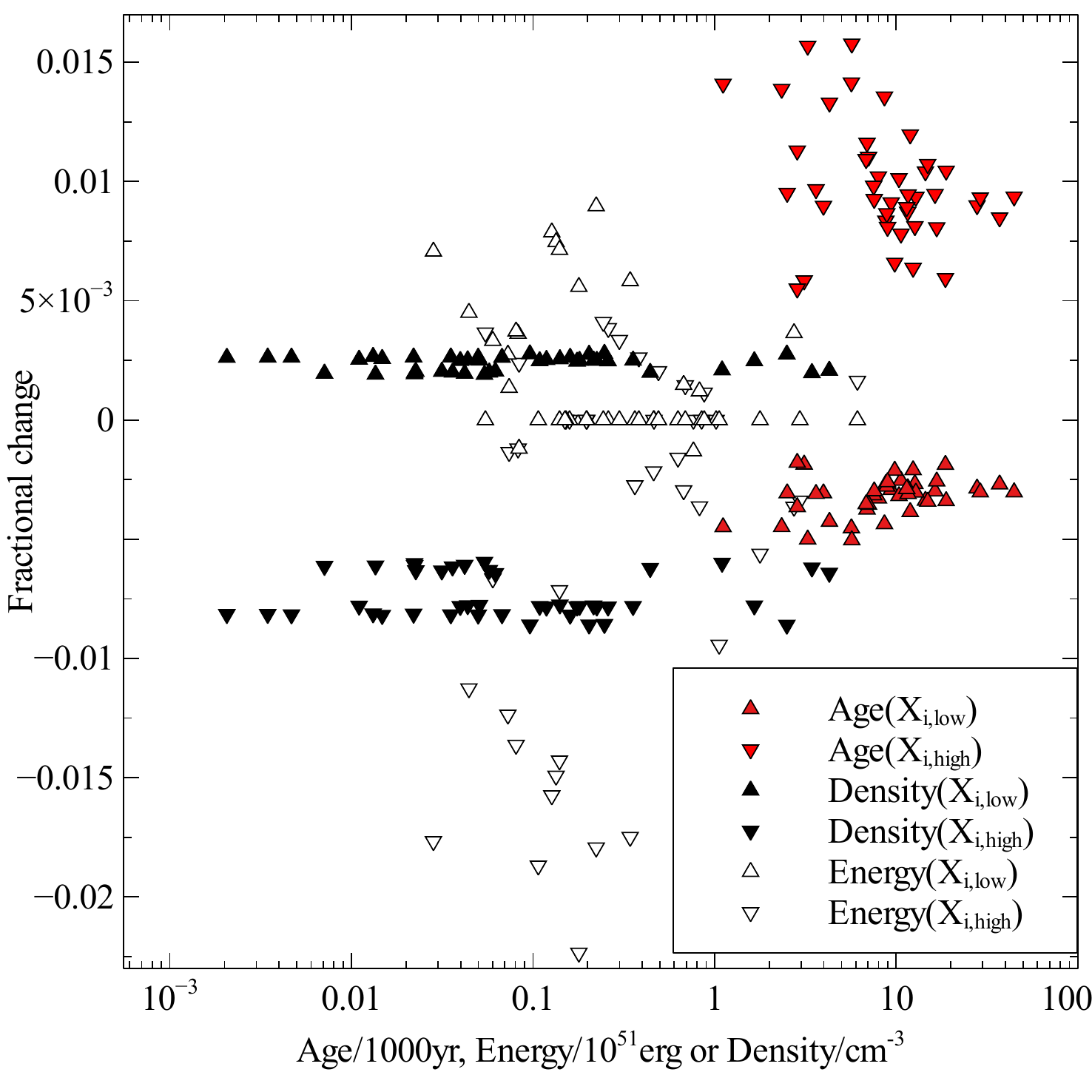}
\caption{Left panel :Plot of energy vs. density for the Standard (s=0, n=7) models of the 43 SNRs. Right panel: Standard model run for the 43 SNRs with ISM abundances higher and lower than standard abundaces, by factor of $10^{0.5}$. 
The fractional change, defined as (new value - standard value)/standard value, is shown for age,
density and explosion energy, and shown for the high abundance case (labelled $X_{i,high}$, downward 
pointing triangles) and for the low abundance case (labelled $X_{i,low}$, upward 
pointing triangles). All output values change less than 2\% as ISM abundance is changed.
}
\label{fig:EvD_ISMabund}
\end{figure*}

We examine some of the systematic effects that could be in the models, including sensitivity of model results 
to some of the changes in input assumptions. 
No correlation is expected between explosion energies and interstellar densities.
The plot is shown in the left panel of Figure~\ref{fig:EvD_ISMabund}, using the Standard model results from Table~\ref{tab:TBLstdmodel}. 
A least squares regression fit to energy vs. density yields a nearly flat line 
with a Pearson correlation coefficient of  $R^2=0.0073$, thus no correlation is seen between the two quantities.

Next we consider the effect of changing the ISM abundances. The Standard model is run
with a high abundance case with all elements except H and He elevated in abundance by 
a factor of $10^{0.5}$, and a a high abundance case with abundances decreased by the same factor 
compared to the standard solar ISM abundance.
The results are shown in the right panel of  Figure~\ref{fig:EvD_ISMabund}.
The fractional change, defined as (new value - standard value)/standard value, is shown for age,
density and explosion energy.
For  age, the high abundance case yields a $\sim 1.2$\% increase, and low abundance yields  
a $\sim 0.4$\% decrease.
For density, the change is systematic with high abundance yielding a $-0.7$\% change in ISM density and
low abundance yielding a $+0.25$\% change.
For  explosion energy, the high abundance case yields a $\sim 1.5$\% decrease, and low abundance yields  
a $\sim 0$ to $\sim0.8$\% increase.
All output values change less than 2\% as ISM abundance is changed.

\begin{figure*}[ht!]
\plottwo{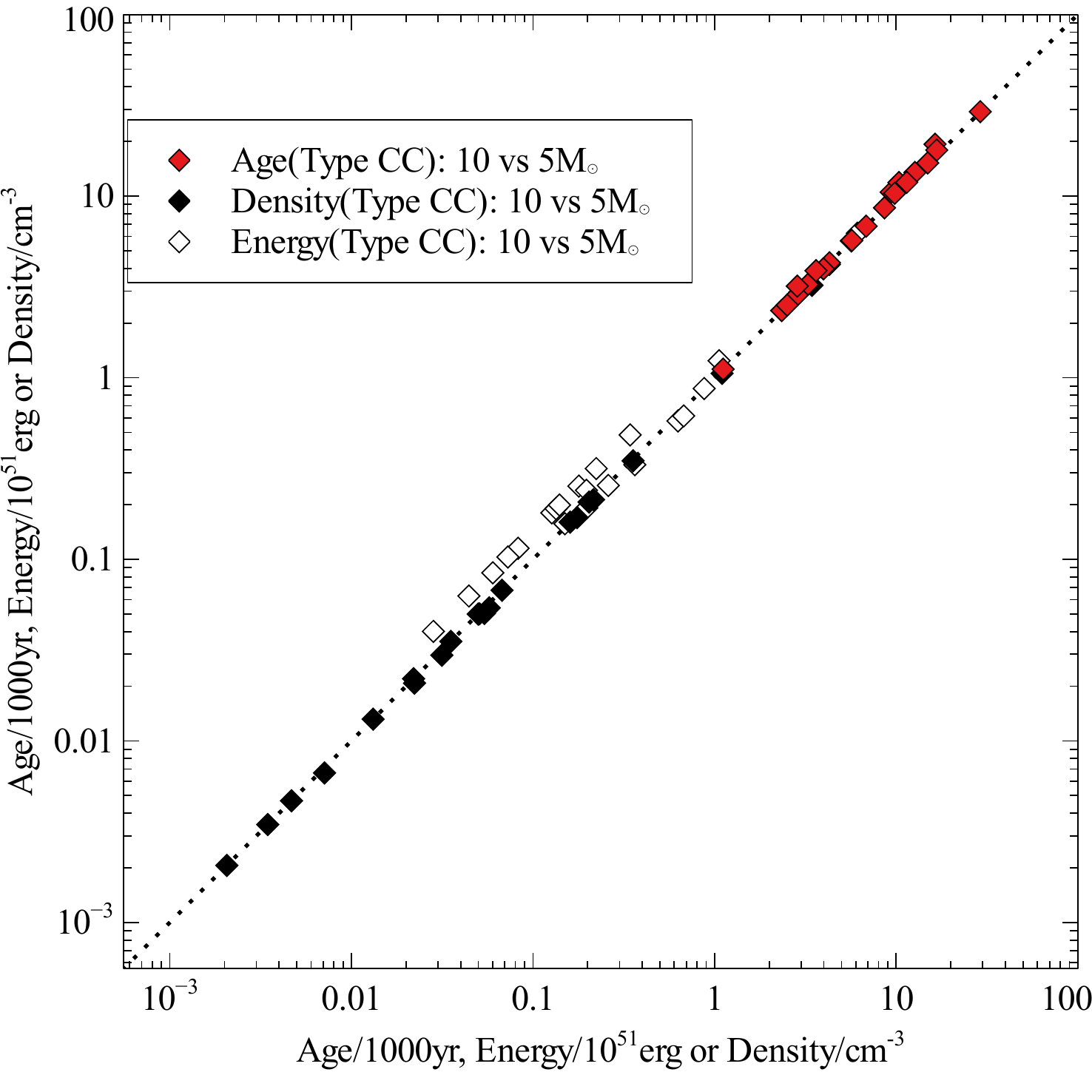}{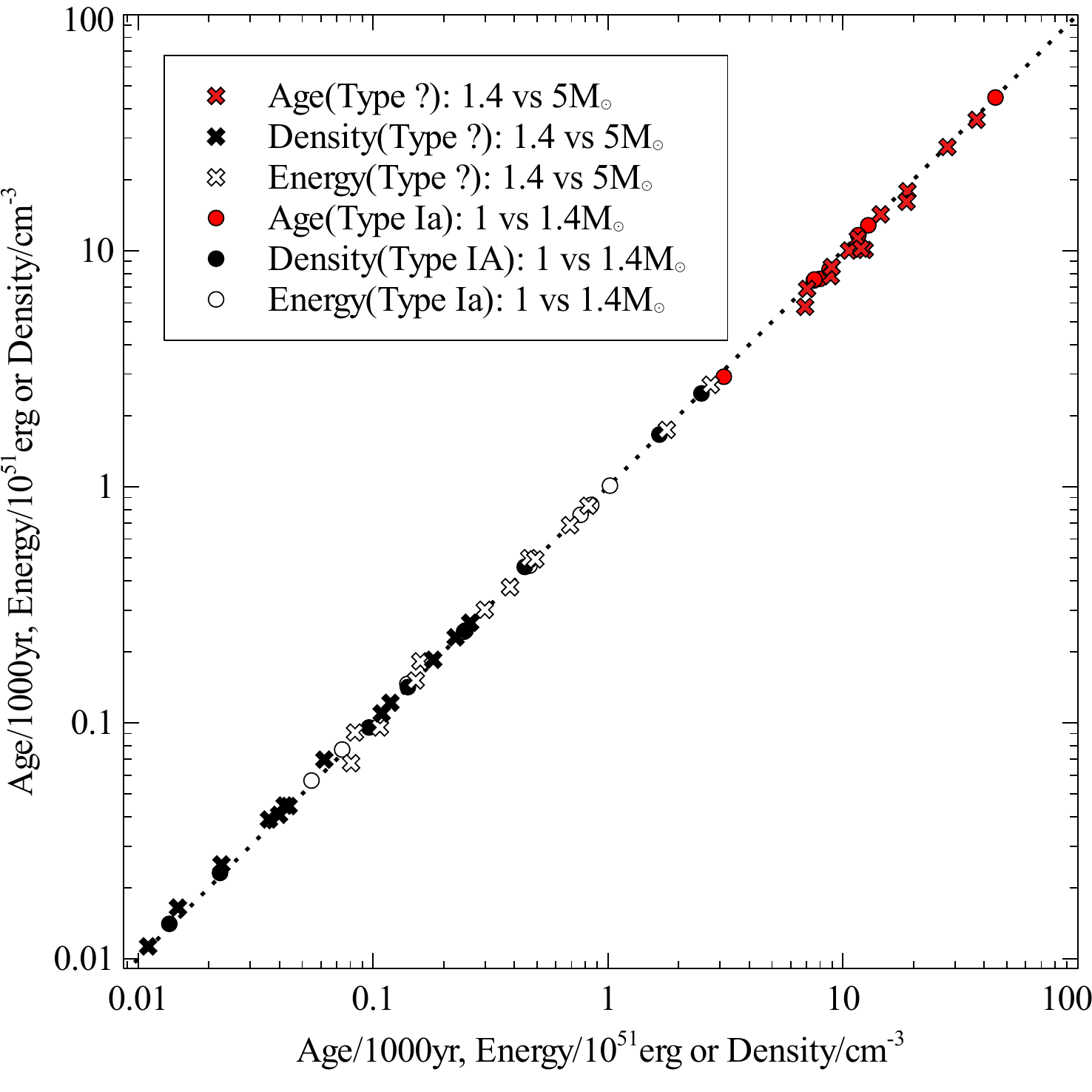}
\plottwo{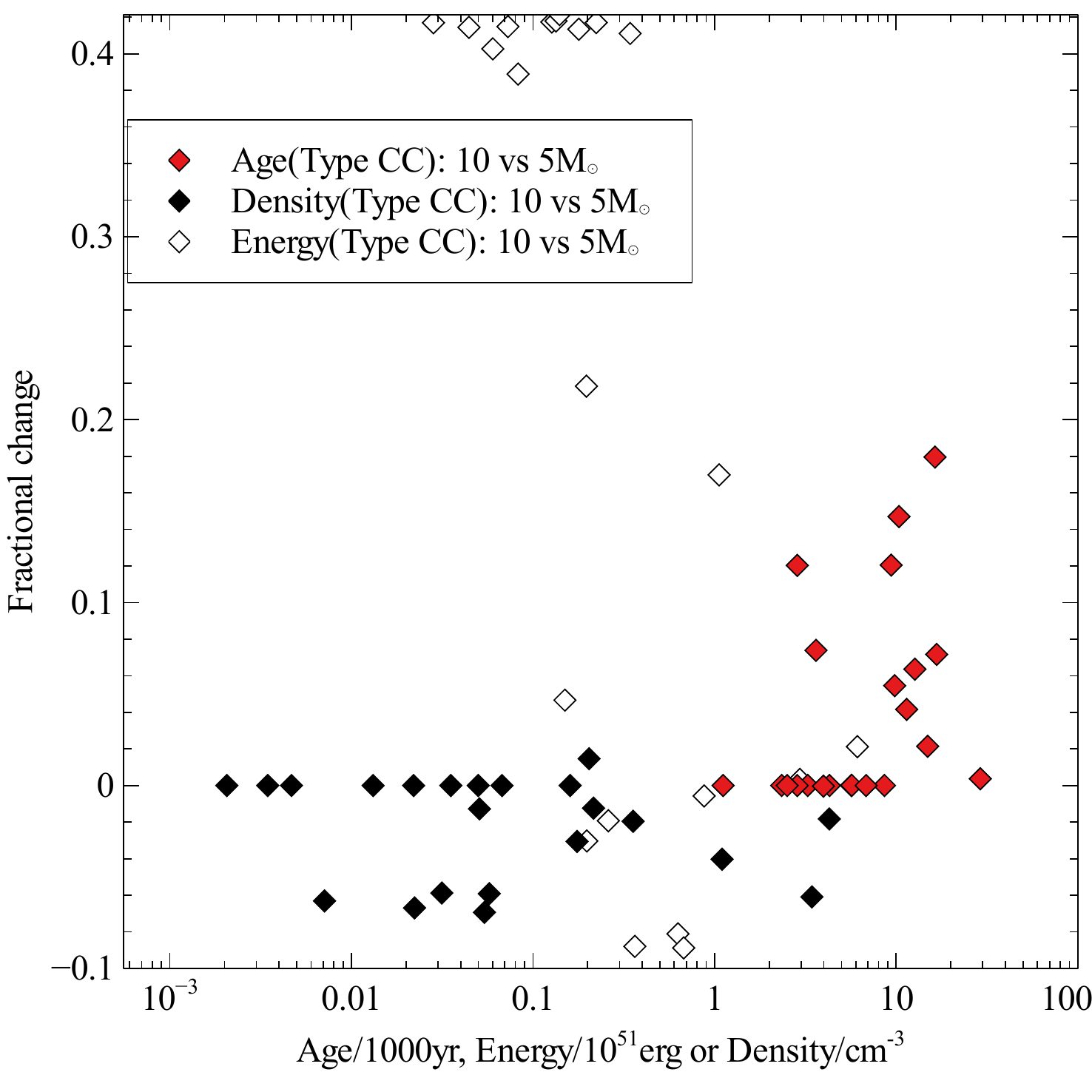}{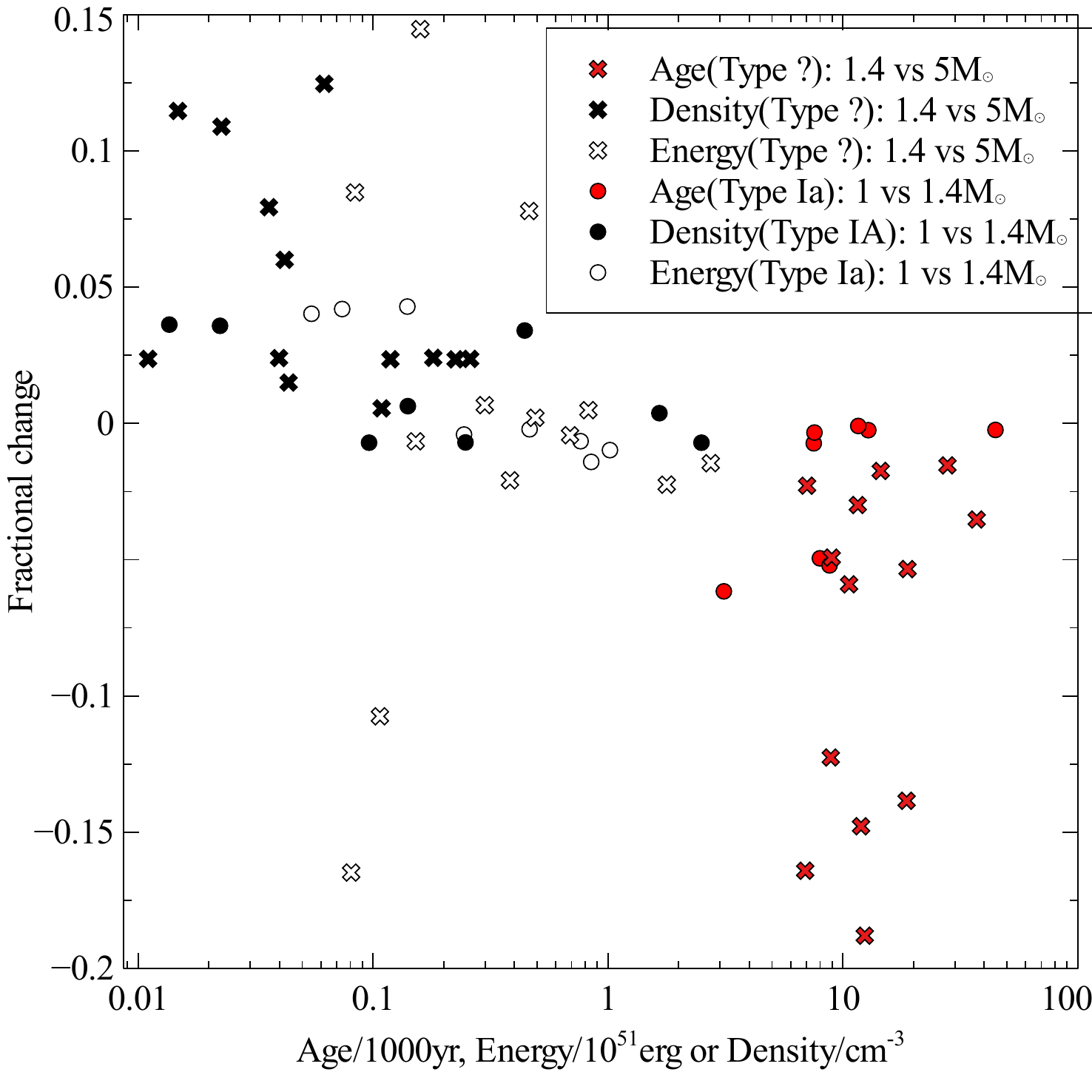}
\caption{The effect of changing ejecta mass on the model outputs: 
The upper left panel is for Type CC SNRs: the age,
density or energy with changed mass vs age, density or energy with standard mass.
The upper right panel is for Type ? and Type Ia SNRs: the age,
density or energy with changed mass vs age, density or energy with standard mass.
 (Lower left panel) the fractional change in age vs. age for the subset of 23 Type CC SNRs;
the fractional change in density vs. density for Type CC; and the fractional change in energy vs. energy
for Type CC. 
 (Lower right panel) the fractional change in age vs. age for the subset of 13 unknown type (Type ?) SNRs
 and for the 8 Type Ia SNRs;
the fractional change in density vs. density for Type ? and Type Ia; and the fractional change in energy vs. energy for Type ? and Type Ia.
}
\label{fig:ejectamass}
\end{figure*}

The effect of changing the assumed ejecta masses is considered next. 
For this test, the mass for CC Type SNRs is increased from 5 to 10 $M_{\odot}$,
for Type Ia SNRs is decreased from 1.4 to 1.0 $M_{\odot}$. For unknown types (Type ?),
we change the mass from 5 to 1.4 $M_{\odot}$. The reason for the last change is to 
estimate the effect of having the Type ? as Type Ia instead of Type CC.
The comparison of new age, density and explosion energy with the Standard model values is shown in 
the top panels of Figure~\ref{fig:ejectamass}. 
Type CC are on the left and Type ? and Type Ia are on the right.
On these plots, the change in age, density or energy is small compared with their values 
so the values lie close to the line y=x.
To examine the effect of changing ejecta mass more closely,
the fractional changes for age, density and explosion energy are shown in the lower panels 
of Figure~\ref{fig:ejectamass}. Type CC are on the left and Type ? and Type Ia are on the right.

For Type CC SNRs, increasing ejecta mass from 5 to 10$M_{\odot}$ increases the age by 0 
to 20\%, with mean change of 4\%.
Increased ejecta mass decreases the ISM density by 0 to 7\%, with mean decrease of 2\%.
The explosion energy has the largest change, from a 9\% decrease to a 42\% increase, with mean increase of 18\% and large standard deviation of 20\%.
For Type Ia SNRs, decreasing ejecta mass from 1.4 to 1$M_{\odot}$ decreases the age by 0 
to 6\%, with mean decrease of 2\%.
The decreased ejecta mass changes the ISM density by -1 to +3.5\%, with mean increase of 1.5\%.
The explosion energy for Type Ia has a change of  -1 to +4.5\% increase, with mean increase of 2\%.
For Type ? SNRs, decreasing ejecta mass from 5 to 1.4$M_{\odot}$ effectively assigns them as 
Type Ia instead of Type CC. This has the following effects. Age decreases  by 2  
to 19\%, with mean decrease of 7\%.
The ISM density increases by 0 to 13\%, with mean increase of 4\%.
The explosion energy has the largest change, from a 17\% decrease to a 14\% increase, with mean increase of 1\%. 

Overall, the change in mean age and mean ISM density for all three cases is small (between 1\% and 4\%),
 whereas the change in mean explosion energy is moderate (between 1\% and 20\%).
 The effect of changing ejecta mass has a negligible effect on the age and density distributions for 
 our sample, shown in  Figure~\ref{fig:age_energy}. 
 It has a small effect on the energy distribution.
 The energy distribution does not significantly change, and the derived parameters of the log-normal fit to explosion energies change to $E_{av}= 3.0\times10^{50}$ erg and dispersion $\sigma_{logE}$=0.53.

The different possible choices of SNR type, via choice of s and n, affect the output results of the model.
This is discussed in detail in \citet{2019AJ....158..149L} and in  \citet{2017LeahyWilliams}. 
 For example, 8-9 different models are applied to each of three historical SNRs (Table 4 of \citealt{2019AJ....158..149L}).
 That work demonstrated that s=2 models are $\sim$5-10 times younger and their explosion energies are $\sim$10-100 times larger than for s=0 models.
 The ratio of emission measure of shocked ejecta to that of shocked ISM is much larger for s=2 models than for s=0.
 This large difference between s=0 and s=2 models allows one to distinguish between the two cases. 
 For historical SNRs, the known age allows distinction. For SNRs of unknown age, one requires another quantity than age:
 a very useful one is the ratio of measured  EM for shocked ejecta to EM for shocked ISM.
In section~\ref{sec:2comp} below we use SNRs with measured shocked ISM and shocked ejecta  to distinguish whether the type is s=0 or s=2. 

\subsection{Cloudy ISM SNR Model}\label{sec:cloudy}

\begin{figure*}[ht!]
\includegraphics[width=\textwidth]{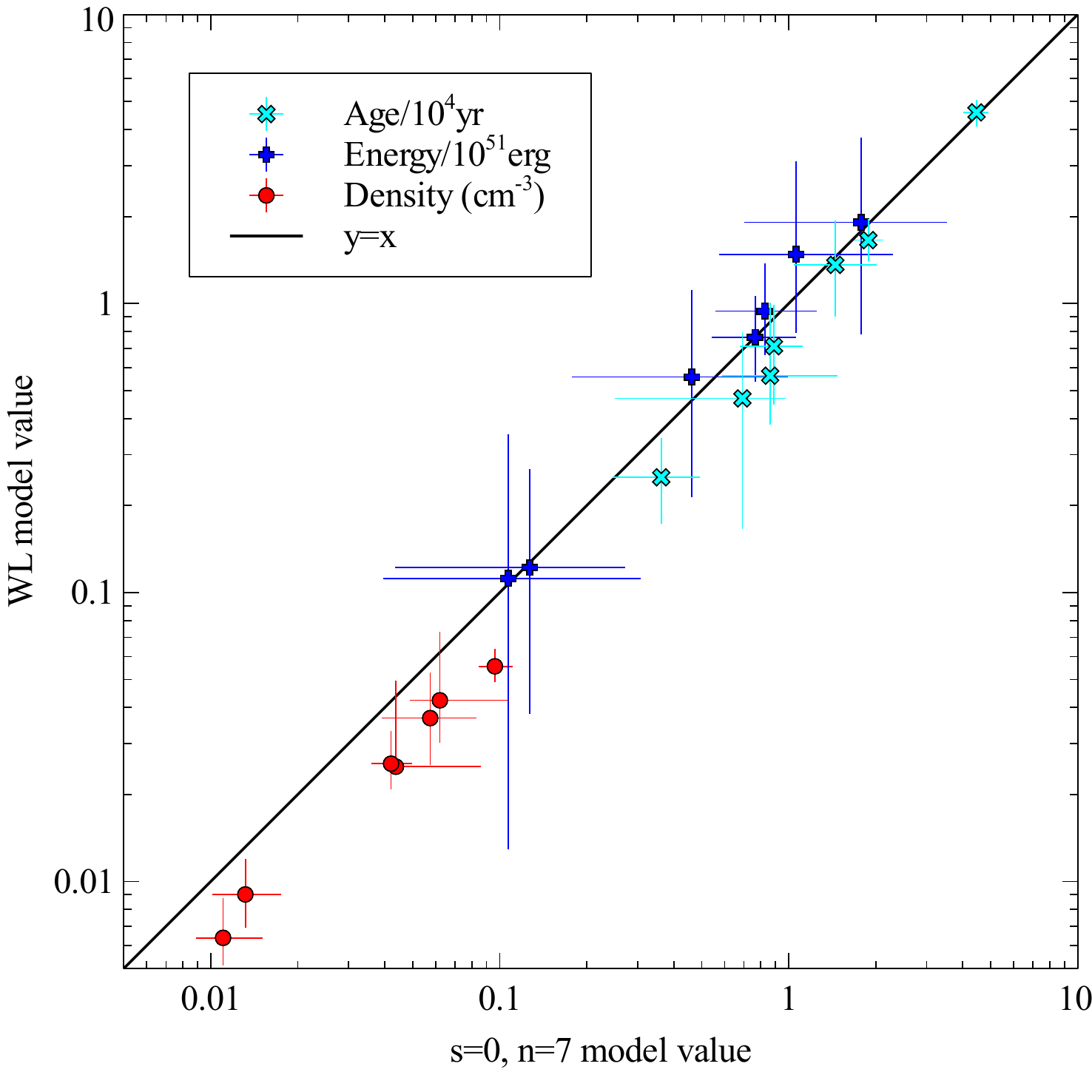}
\caption{Comparison of ages, energies and densities of the 7 mixed morphology SNRs from the s=0, n=7 model and 
from the cloudy SNR (WL) model. 
}
\label{fig:WLvs0n7}
\end{figure*}

The resulting age, energy and density parameters are compared to those from the standard s=0, n=7 model in
Figure~\ref{fig:WLvs0n7}.
The ages and explosion energies are the same within errors for all 7 SNRs.
The ISM density is lower for the WL model than for the s=0, n=7 model by a factor of $\simeq1.5$.
A lower derived ISM density for the WL model is expected because the WL model contains clouds.
The clouds evaporate over the age of the SNR to increase the current post-shock density compared to what it would
be without evaporation.
This means that the initial ISM intercloud density is lower for the WL model than it would be for a standard uniform
density ISM model.

\subsection{Two component SNRs}\label{sec:2comp}

We compare the  predicted  $EM_{RS}$ and $kT_{RS}$ of the different (s,n) cases with the measured values 
for each of the 12 SNRs to select a preferred model for each SNR.
In choosing a preferred model, we attach more importance to reproducing $EM_{RS}$ and less to $kT_{RS}$.
There are two reasons: i) uncertainties in the electron ion equilibration process for the reverse shock; 
ii)  $kT_{RS}$ is more strongly affected than $EM_{RS}$ by the abundances in the ejecta \citep{2018ApJ...866....9L}. 
Here, we assume fixed standard CC or Ia abundances rather than fine-tuning them. Instead we allow
for differences of a factor of a few between the predicted and measured $EM_{RS}$ to indicate agreement.
This is reasonable considering the several orders of magnitude difference in predicted $EM_{RS}$ from
the different cases of (s,n) that we calculate.

The SNRs (with SN types labelled) which have the (s,n)=(0,7) model yield $EM_{RS}$ closest to the measured values
are G116.9+0.2 (CC),  G332.4-0.4 (CC?) and G350.1-0.3 (CC).
None of the 12 SNRs have the  (s,n)=(0,10) model yield the closest agreement.
The SNRs which have the (s,n)=(2,7) model yield $EM_{RS}$ closest to the measured values
are G109.1-1.0 (CC),  G156.2+5.7 (CC), G330.0+15.0(?), and  G349.7+0.2(CC).
The SNRs which have the (s,n)=(2,10) model yield $EM_{RS}$  closest to the measured values
are G53.6-2.2 (Ia),  G299.2-2.9 (Ia), G306.3-0.9 (Ia),  G315.4-2.3 (Ia) and  G352.7-0.1 (Ia?).

\begin{figure*}[ht!]
\includegraphics[width=\textwidth]{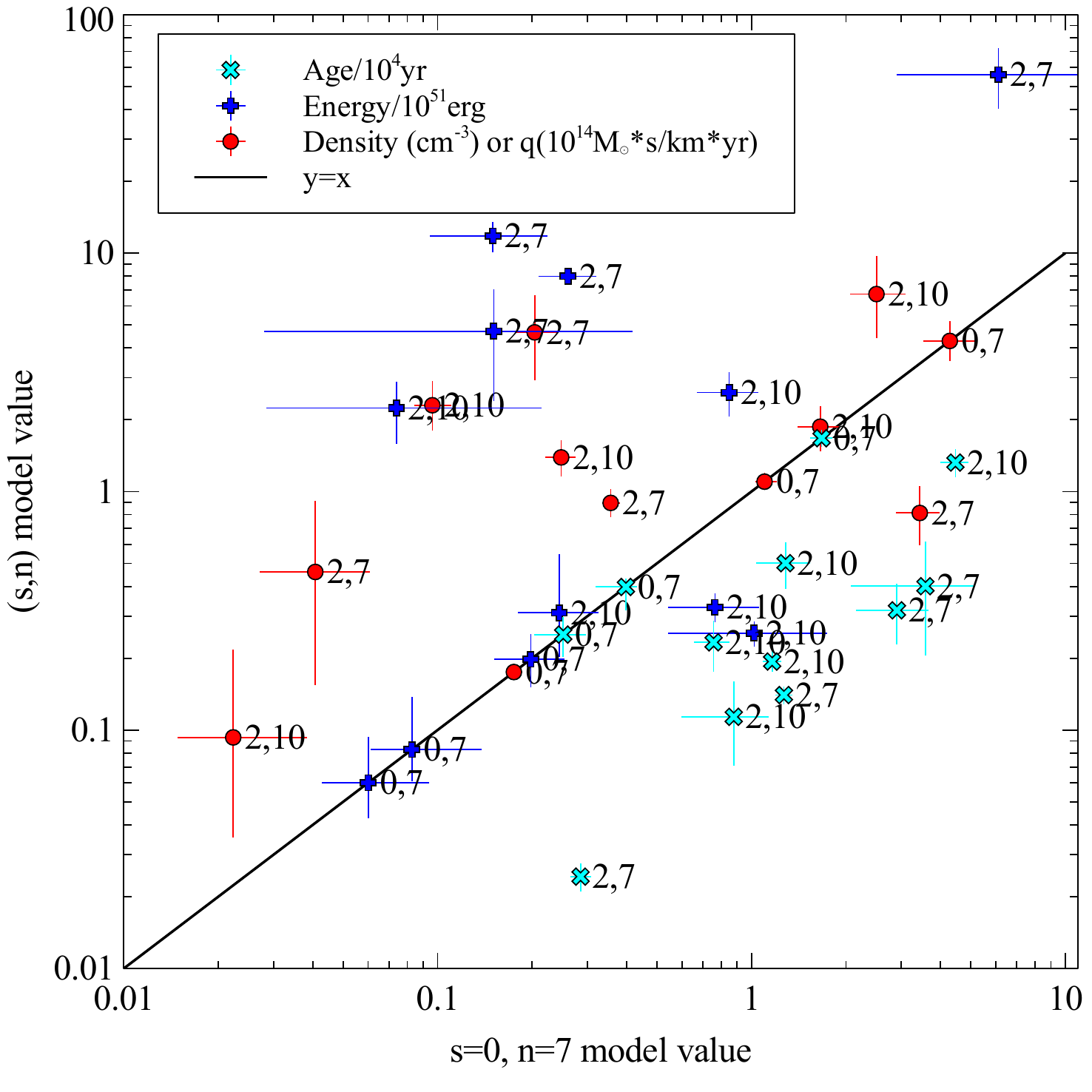}
\caption{Ages, energies and densities of the preferred (s,n) models for each of 12 SNRs with two emission components. 
These are compared to the standard (s=0, n=7) model for each SNR. 
Each point is labelled with the preferred value of (s,n). 
 In most cases, s=2 models work better than s=0 models. 
 On average, the energies of the s=2 models are higher than the (0,7) model and the ages are smaller. 
}
\label{fig:customvs0n7}
\end{figure*}

Figure~\ref{fig:customvs0n7} shows the results of our best (s,n) model for each of the 12 SNRs 
compared to the values from the standard (0,7) model. The ages and energies are directly compared.
The ISM density, $n_0$ and the stellar wind parameter $q=\dot M/(4 \pi V_w)$ are plotted to show the wide range
of $n_0$ (0.02 to 4 cm$^{-3}$) and $q$ ($10^{13}$ to $7\times10^{14}$), although the $y=x$ line is not
applicable.

Allowing for different s and n for each SNR makes a significant difference. 
We find that 9 of the 12 SNRs are more consistent with a stellar wind SN than an SN in a uniform ISM. 
This makes a large difference in the SN energy and SN age. 
For the 12 SNR (0,7) models, the mean log(age) and its standard deviation are 4.04 and 0.41, which
decrease to 3.45 and 0.48, when individual customized (s,n) models are used. 
This is an average decrease by factor 3.9 in age.
The mean log(energy) and standard deviation are 50.47 and 0.58, which
increase to 51.05 and 0.94, when individual customized (s,n) models are used.
This is an average increase by factor 3.8 in energy.
A least squares fit of a log normal distribution to the energy values of the 12 SNRs yields a good fit with 
a mean log($E$) of 51.01 and $\sigma(log(E))$ of 1.08, consistent with the above values.
These changes are large, and point to the importance of applying the correct type of SNR
model when interpreting the measured values of $kT$ and $EM$. 
In particular, the statistics inferred for a SNR population when using only a s=0, n=7 model,
may be offset to larger ages and lower explosion energies than using better (general s and n) models.

\subsubsection{Type Ia SNRs with two component spectra}\label{sec:typeIa}

An important result that emerges from the models for two component SNRs, is that all five of the
type Ia SNRs are consistent with stellar wind explosions and not with uniform density ISM explosions.
This indicates that the origin of at least these five type Ia must be associated with a stellar wind.
The stellar wind could be from a single WD accreting from a companion, or it could be binary WD merger
if the merger occurs before the wind dissipates. 

Previous work has found evidence for type Ia SNRs to be associated with stellar wind bubbles or cavities. \citet{1999ApJ...527..866L} measured the HI environment of the probable type Ia SNR G93.3+6.9/DA530 and found it lies within a shell likely created by a stellar wind.
\citet{2012ApJ...756....6P} study Kepler's SNR with hydrodynamical and X-ray spectral modeling. 
They find neither uniform density nor pure $1/r^2$ wind profile fits but an explosion in a small cavity, 
indicative of some type of wind, gives consistency with observations.
\citet{2014Broersen} use hydrodynamic simulations to find that G315.4-2.3/RCW 86 is a Type Ia 
explosion in a stellar wind cavity, as we find here.
\citet{2016ApJ...826...34Z} study Tycho's SNR using molecular line observations to detect an expanding 
clumpy bubble at the boundary of the SNR, likely the fast wind from an accreting WD progenitor.

The work here shows that to reproduce the measured ejecta $EM_{RS}$ and ISM $EM_{FS}$,
 s=0 models are ruled out and s=2 models favored for the type Ia SNRs.
The current modelling of SNRs assumes spherical symmetry as an approximation to real SNRs. 
It is possible that more complex geometries, which do not require a stellar wind, could give
 give high ratio of ejecta EM to shocked-ISM EM. 
 I.e., they could fit both measured $EM_{RS}$ and $EM_{FS}$ for type Ia SNR, like the s=2 models  
 and unlike the s=0 models.

\subsection{SNR birthrate and energetics}\label{sec:2comp}

Most X-ray spectrum observations of SNRs do not yield two emission components, from forward shock and from ejecta.
This can result from observations that are not sensitive enough, or
if the SNR is old enough for the ejecta component to have faded below detection limits. 
For these SNRs we cannot constrain which (s,n) model fits better.  

For the current set of 43 SNRs, we estimate the effect of having a subset of the SNRs in a stellar 
wind (s=2) rather than a uniform ISM (s=0) in two ways.
First we use all of the SNRs from the current work and from \citet{2018ApJ...866....9L}.
The 43 SNRs consists of 34 with s=0 and 9 with s=2.
The 15 SNRs of \citet{2018ApJ...866....9L} (Table 4 in that paper) had 13 with s=0 preferred
and 2 with s=2 preferred. 
Thus, 11 of the 58 SNR models have a stellar wind environment.
 
Because a second EM is needed to decide if a s=2 model fits better than an s=0 model, 
the s=2 SNRs are all from SNRs with 2 measured EM components. 
Thus the above number gives a lower limit on the fraction of stellar wind SNRs.
Using only 2 component SNRs, 2 of 5 SNRs from \citet{2018ApJ...866....9L} had s=2
and 9 of 12 in the current sample have s=2, giving a fraction of 11 of 17 with s=2.
It is likely that this is an overestimate because s=0 SNRs have smaller ejecta EMs,
thus are less likely to be included in the samples with 2 measured EMs.
 
Overall, we use 11/58 ($0.19\pm0.06$) as the lower limit and 11/17 ($0.65\pm0.20$) as 
the upper limit on fraction of SNRs in a stellar wind environment, including counting statistics error.   
The average of the upper and lower limits gives $0.42\pm0.28$ as the best estimate of fraction of stellar wind SNRs. 
For 43 standard (s=0,n=7)  models, we account for the presence of stellar wind SNRs using 
a fraction $\sim$0.4 which should be s=2 models. From Section~\ref{sec:2comp}, the s=2 ones 
have mean energy higher by a factor of 3.8 and mean age lower by a factor of 3.9 than s=0 models.
This yields the mean energy for the log-normal distribution to be a factor $\sim$2 larger, and   
the mean age of the whole sample to be a factor $\sim$0.7 smaller. 
The estimated corrected explosion energy is 
$5\times10^{50}$ erg and corrected SNR birth rate is 1/250 yr. 

Because our sample is incomplete, a more complete survey of SNRs will increase the birth rate.
We include the 15 Galactic SNRs from the sample of \citet{2018ApJ...866....9L}, 
of which 13 have ages less than 10,000 yrs. 
Then our distribution of ages (similar to Figure~\ref{fig:age_energy} here) has a best-fit birth rate of 1/230 yr. 
If we further correct for expected fraction of stellar wind SNRs, the Galactic birth rate increases to 1/160 yr. 

This rate is still significantly less than the SN rate in the Galaxy of  1 per 40$\pm 10$ yr  \citep{1994Tammann}.
The difference is primarily caused by incompleteness in SNRs measured in X-rays.
Including the 58 Galactic SNRs with distances and X-ray spectra (this work and 
\citealt{2018ApJ...866....9L}) and comparing to the list of 294 SNRs \citep{2019Green},
we estimate an incompleteness factor of $\sim5$\footnote{This is uncertain by a factor $\sim$2: 
the X-ray lifetime of an SNR may be shorter than the radio lifetime, which decreases the correction factor; and radio observations are incomplete, which increases the correction factor.}.
An alternate way to estimate the incompleteness is to consider the distribution of this sample of 
43 SNRs with distance. A plot of cumulative number of SNRs vs. distance, $d$, 
shows a decrease from the expected $d^2$ increase for a disk-like distribution beyond a distance
of $\sim$5 kpc. The area of the Galaxy, using most SNRs occur within $\sim$12 kpc of the center, 
to the area of a 5 kpc circle is $\sim$6, consistent with an incompleteness factor of $\sim$5
for SNRs measured in X-rays.

The net result is that the SNR birth rate corrected for incompleteness increases to $\sim$1 per 35 yrs.
This latter rate is consistent with the estimated SN rate.
This implies that there is no missing SNR problem, with large uncertainties, from statistics of X-ray detected 
SNRs\footnote{There may be a missing young SNR problem. 
In the current study we omit the well-studied historical SNRs (with ages $\lesssim 1000$ yr) and 
instead allow an offset in the fit of birth rate to our distribution of SNR ages.}. 

Regardless of whether the incompleteness correction factor is overestimated or not, 
the ages of a significant fraction Galactic SNRs are overestimated.
Many ages are estimated by using too-simplistic models, such as Sedov models with assumed energies and ISM densities.
\citet{2017LeahyWilliams} showed that Sedov models were significantly offset compared to  
more accurate TM99 models with the same input parameters.
As shown in \citet{2018ApJ...866....9L}, inclusion of $EM$ and $kT$ in modelling SNRs 
affects significantly estimates of ages and explosion energies. 
As shown here, inclusion of both uniform ISM and stellar wind SNR models has the important effect of
increasing energies and decreasing ages compared to using only uniform ISM SNR models.

\section{Conclusion}\label{sec:conclusion}

Distances to Galactic SNRs have improved significantly over the past decade, allowing determination of radii.
X-ray observations of SNRs have been carried out for a significant fraction of Galactic SNRs.
Together, these enable the application of SNR models to observed radii, emission measures and temperatures.
We have improved the accuracy of spherically symmetric SNR evolution models by including results 
from a large grid of hydrodynamic simulations \citep{2019AJ....158..149L}.
In the current study we apply the models to estimate SNR parameters for a sample of 43 Galactic SNRs.
The distributions of the parameters were used to estimate properties of the Galactic SNR population.
The energies and ISM densities of SNRs can be well fit with log-normal distributions in agreement
with our earlier studies (\citealt{2017Leahy}, \citealt{2018ApJ...866....9L}).

Seven of the 43 SNRs are mixed morphology SNRs. For these we apply WL cloudy ISM SNR models and compare the
WL parameters to those estimated using our s=0, n=7 models. We find close agreement in parameters.

12 of the SNRs have two component X-ray spectra, indicating that both forward shocked material and
reverse shocked ejecta are detected. 
With the extra information, we can distinguish between models with different s and n. 
We select the (s,n) model closest to the observations. 
One important finding is that all 5 of the 12 two-component SNRs that are type Ia are consistent with a stellar wind SNR. 
This has important implications for the progenitor types for these 5 type Ia SN: they should be consistent with a stellar wind environment.
 
A second important conclusion is that the inclusion of models for stellar wind SNRs is important when assessing 
 the energies and ages of SNRs. 
We estimate that $\sim$40\% of SNRs are stellar wind type, and the effect this has
on SNR population ages and explosion energies.
Including both uniform ISM and stellar wind SNR models has a
significant effect on inferred mean ages of SNRs (a decrease by factor $\sim$0.7) and on
inferred mean explosion energies of SNRs (an increase by factor of $\sim$2).
 After correction for stellar wind SNRs, we find a mean explosion energy of $\sim5\times10^{50}$ erg and 
 a SNR birth rate of 1/260 yr for our population of 43 SNRs.
Inclusion of the 15 Galactic SNRs from \citet{2018ApJ...866....9L} increases the birth rate to 1/160 yr.
A final correction for the incompleteness of X-ray observations of SNRs increases the SNR birth
rate to  $\sim$1/35 yr, which is consistent with the Galactic  SN rate \citep{1994Tammann}. 
To within uncertainties, this implies that there is no missing SNR problem when considering X-ray data. 

This work was focussed on the population properties of Galactic SNRs by analyzing measured $EM$ and 
$kT$ from X-ray observations using of a basic set of SNR models. 
Future work will focus on individual SNRs where detailed consideration is given to observations
at other wavebands to constrain better the choice SNR model (e.g., (s,n) parameters, ejecta mass
and abundances).
 
\acknowledgments
This work was supported by a grant from the Natural Sciences and Engineering Research Council of Canada.
The authors thank the referee and editor for providing useful comments which improved this work.

\clearpage



\clearpage

\begin{deluxetable}{crrrrrrr}
\tabletypesize{\scriptsize}
\tablecaption{Observed Quantities of 43 SNRs} 
\tablewidth{0pt}
\tablehead{
\colhead{SNR} & \colhead{Type}  & \colhead{Distance} & \colhead{Semi-axes}  &  \colhead{Radius}      & \colhead {$EM^{a,b}$}     & \colhead {$kT^{a,b}$} & \colhead{Refs$^{f}$}\\
\colhead{}      & \colhead{}           & \colhead{(kpc)}   & \colhead{(arcmin)}   & \colhead{(pc)}      & \colhead {($10^{58}$cm$^{-3}$)} & \colhead {(keV)} &  \\
}
\startdata
G38.7-1.3  &  CC? & 4$\pm0.8^{c}$  & 16$\times$9.5  & 14.9$\pm3.0^{d}$& 0.024(0.017-0.031)d$_{4}^2$        & 0.65(0.35-0.95)   & 1, 1, 1 \\   
G53.6-2.2  & Ia & 7.8$\pm$0.8    & 16.5$\times$14 & 34.6$\pm$3.6    & 11.4(9.7-13.4)d$_{7.8}^2$         & 0.14(0.13-0.15)   & 2, (3,4), 2 \\
           &    &                &                &                 & 0.022(0.015-0.030)d$_{7.8}^2$   & 3.9(3.6-4.2)      & -, -, 5 \\  
G67.7+1.8  &  CC? & 2.0$^{+3.7}_{-0.5}$& 7.5$\times$6   & 4$^{+7}_{-1}$    & 0.0033(0.0015-0.0056)d$_{2}^2$          & 0.59(0.53-0.65)   & 6, 4, 6 \\ 
G78.2+2.1  &  CC? & 2.1$\pm0.4$    & 30$\times$30   & 19$\pm$4        & 0.61(0.30-1.0)d$_{2.1}^2$          & 0.83(0.62-1.0)    & 7, 7, 7 \\
G82.2+5.3  &  ? & 3.2$\pm0.4$    & 47$\times$32   & 37$\pm$4        & 0.184(0.132-.308)d$_{3.2}^2$       & 0.63(0.59-0.68)   & -, 4, 8 \\ 
G84.2-0.8  &  ? & 6.0$\pm$0.2    & 10$\times$8    & 18$\pm$3        & 0.11(0.08-0.13)d$_{6}^2$           & 0.58(0.49-0.66)   & -, 9, 9 \\ 
G85.4+0.7  & CC?& 3.5$\pm$1      & 12$\times$12   & 12$\pm$4        & 0.035(0.027-0.061)d$_{3.5}^2$      & 0.86(0.81-0.90)   & 10, 11, 11 \\
G85.9-0.6  & Ia?& 4.8$\pm$1.6    & 12$\times$12   & 17$\pm$5        & 0.030(.027-.036)d$_{4.8}^2$        & 0.68(0.64-0.71)   & 10, 11, 11 \\ 
G89.0+4.7  & CC & 1.9$^{+0.3}_{-0.2}$  & 60$\times$45  & 31$\pm$4   & 0.051(.043-.064)d$_{1.9}^2$        & 0.61(0.58-0.65)   & 12, 4, 13 \\ 
G109.1-1.0 & CC & 3.1$\pm0.2$    & 14$\times$14   & 12.6$\pm$0.9    & 7.5(7.1-8.0)d$_{3.1}^2$            & 0.255(0.245-0.265)& 14, 15, 16 \\ 
           &    &                &                &                 & 1.13(1.05-1.2)d$_{3.1}^2$          & 0.64(0.63-0.68)   & -, -,  17 \\  
G116.9+0.2 & CC & 3.1$\pm$0.3    & 17$\times$17   &  15.3$\pm$1.5   & 3.26(3.21-3.31)d$_{3.1}^2$         & 0.22(0.19-0.25)   & 18, 19, 13\\
           &    &                &                &                 & 0.0115(0.011-0.012)d$_{3.1}^2$     & 0.84(0.78-0.93)   & -, -, 20\\   
G132.7+1.3 & ?  & 2.1$\pm$0.1    & 40$\times$40   &  24$\pm$1.2     & 4.72(4.67-4.77)d$_{2.1}^2$         & 0.22(0.17-0.28)   & -, 21, 13 \\ 
G156.2+5.7 & CC & 2.5$\pm0.8$    & 55$\times$55   &  38$\pm$10      & 67(66-68)d$_{2.5}^2$               & 0.40(0.39-0.41)   & 22, 23, 22\\
           &    &                &                &                 & 3.4(3.3-3.5)d$_{2.5}^2$       & 0.51(0.50-0.52)   & -, -, 22 \\   
G160.9+2.6 & CC?& 0.8$\pm$0.4    & 70$\times$60   &  16$\pm$8       & 0.0038(0.0034-0.0042)d$_{0.8}^2$   & 0.82(0.76-0.88)   & 12, 24, 25 \\ 
G166.0+4.3 & ?  & 4.5$\pm$1.5    & 27$\times$18   &  30$\pm$9       & 1.5(1.2-1.8)d$_{4.5}^2$            & 0.8(0.5-1.1)      & -, 26, 27 \\ 
G260.4-3.4 & CC & 1.3$\pm$0.3    & 30$\times$25   &  10.5$\pm$3     & 0.00058(0.00046-0.00070)d$_{1.3}^2$& 0.79(0.66-0.92)   & 28, 29, 30 \\ 
G272.2-3.2 & Ia & 6$\pm4$        & 7.5$\times$7.5 &  14$\pm$7       & 1.57(1.49-1.65)d$_{6}^2$           & 0.73(0.68-0.76)   & 31, 31, 32 \\ 
G296.7-0.9 & ?  & 10$\pm0.9$     & 7.5$\times$4   &  17$\pm$2       & 4.7(3.5-6.7)d$_{10}^2$             & 0.53(0.46-0.60)   & -, 33, 33 \\ 
G296.8-0.3 & CC?  & 9.6$\pm$0.6    & 10$\times$7    &  23.7$\pm$1.5   & 0.21(0.17-0.25)d$_{9.6}^2$         & 0.86(0.85-0.87)   & 35, 34, 35  \\ 
G299.2-2.9 & Ia & 5$\pm1^{c}$    & 9$\times$5.5   &  14$\pm$5       & 0.046(0.025-0.10)d$_{5}^2$         & 0.54(0.42-0.66)   & 36, 36, 37\\
           &    &                &                &                 & 0.029(0.018-0.040)d$_{5}^2$        & 1.36(1.28-1.45)   & -, - ,37 \\   
G304.6+0.1 & ?  & $>9.7^{e}$     & 4$\times$4     &  18$\pm$6       & 0.34(0.30-0.38)d$_{15}^2$          & 0.84(0.75-0.95)   & -, 38, 39\\ 
G306.3-0.9 & Ia & 20$\pm4^{c}$   & 2$\times$2     &  11.6$\pm$2.3   & 290(240-350)d$_{20}^2$                 & 0.190(0.185-0.198)& 40, 41, 42\\
           &    &                &                &                 & 1.8(1.4-2.2)d$_{20}^2$           & 1.51(1.31-1.71)   & -, -, 42 \\ 
G308.4-1.4 & CC?& 3.1$\pm0.3$    & 6$\times$3     &  4.1$\pm$0.5    & 0.074(0.062-0.085)d$_{3.1}^2$         & 0.68(0.52-0.87)   & 43, 44, 45 \\
G309.2-0.6 & CC?& 2.8$\pm0.8$    & 7.5$\times$6   &  8$\pm$4        & 9.E-5(1.E-5-2.0E-4)d$_{2.8}^2$   & 2.0(1.4-3.0)      & , 44, 46\\
G311.5-0.3 & ?  & $>6.6^{e}$     & 2.5$\times$2.5 &  10$\pm$6       & 0.16(0.13-0.19)d$_{13}^2$          & 0.68(0.44-0.88)   & -,38, 47 \\
G315.4-2.3 & Ia & 2.8$\pm0.2$    & 21$\times$21   & 17$\pm$1.1      & 8.5(7.2-9.8)d$_{2.8}^2$            & 0.5(0.44-0.56)    & 48, 49, 48 \\
 &    &     &                &                 & 0.046(0.037-0.056)d$_{2.8}^2$           &  3.04(2.85-3.23)   & , , 48 \\
G322.1+0.0 & CC & 9.3$\pm0.9$    & 4$\times$2.3   & 8.5$\pm$0.8     & 0.064(0.036-0.12)d$_{9.3}^2$       & 0.9(0.5-1.7)      & 50, 50, 51\\
G327.4+0.4 & ?  & 4.3$\pm0.8^{c}$& 12$\times$11   & 13.1$\pm$2.7    & 3.4(2.6-4.2)d$_{4.3}^2$            & 0.41(0.37-0.44)   & -, 52, 53 \\
G330.0+15.0& ?  & 1$\pm0.5$      & 90$\times$90   &  27$\pm$12      & 0.92(0.6-1.2)d$_{1}^2$             & 0.135(0.13-0.14)  & -, 54, 55\\
           &    &                &                &                 & 0.15(0.08-0.22)d$_{1}^2$           &  0.53(0.49-0.57)   & -, -, 55 \\   
G330.2+1.0 & CC  & $>4.9^{e}$     & 5.5$\times$5.5 & 16$\pm$8        & 0.15(0.05-0.4)d$_{10}^2$           & 0.7(0.4-2.0)      & 57, 56, 57 \\
G332.4-0.4 & CC & 3.0$\pm0.3^{c}$& 5$\times$5     &  4.5$\pm$0.9    & 4.25(4.12-4.40)d$_{3}^2$            & 0.57(0.53-0.61)   & 58, 44, 58 \\
           &    &                &                &                 & 0.51(0.47-0.54)d$_{3}^2$         & 0.66(0.59-0.74)   & -, -, 58\\ 
G332.4+0.1 & ?  & 9.2$\pm1.7$    & 7.5$\times$7.5 & 20$\pm$4        & 3.3(2.4-4.2)d$_{9.2}^2$            & 1.34(1.2-1.48)    & -, 59, 59 \\
G332.5-5.6 & ?  & 3.0$\pm0.8 $   & 18$\times$17   & 15.5$\pm$4      & 0.034(0.026-0.042)d$_{3}^2$        & 0.49(0.43-0.57)   & -, 60, 60 \\
G337.2-0.7 & Ia?& 5.5$\pm3.5$    & 3$\times$3     & 4.7$\pm$3       & 0.72(0.54-0.80)d$_{5.5}^2$         & 0.74(0.73-0.77)   & 61, 62, 62 \\
G337.8-0.1 & CC? & 12.3$\pm1.2$   & 4.5$\times$3   & 13.4$\pm$1.3    & 0.29(0.15-0.54)d$_{12.3}^2$        & 1.8(2.6-1.3)      & 63, 64, 63 \\ 
G347.3-0.5 & CC & 1$\pm0.2^{c}$  & 32$\times$28   & 8.8$\pm$1.7     & 0.13(0.03-0.60)d$_{1}^2$           & 0.58(0.51-0.59)   & 65, 66, 65 \\
G348.5+0.1 & CC & 7.9$\pm1.6$    & 8$\times$7     & 17.2$\pm$3.5    & 6.9(4.5-9.0)d$_{7.9}^2$            & 0.55(0.44-0.72)   & 67, 67, 39\\
G348.7+0.3 & CC & 13.2$\pm1.3$   & 9$\times$8     & 32.8$\pm$3.3    & 2.6(1.9-3.8)d$_{13.2}^2$           & 0.89(0.84-0.93)   & 67, 67, 68\\
G349.7+0.2 & CC & 11.5$\pm1.2$   & 1.3$\times$1.0 & 3.7$\pm$0.4     & 21(16-26)d$_{11.5}^2$              & 0.60(0.56-0.64)   & 61, 69, 70 \\
           &    &                &                &                 & 14(12.5-15.5)d$_{11.5}^2$         & 1.24(1.21-1.27)   & -, -,70\\ 
G350.1-0.3 & CC  & 9$\pm1.8^{c}$ & 2$\times$2     & 2.6$\pm$0.5     & 13(10-16)d$_{9}^2$           & 0.48(0.44-0.52)   & 61, 70, 70 \\
           &    &                &                &                 & 1.2(1.0-1.4)d$_{9}^2$         & 1.51(1.42-1.60)   & -, -,70\\ 
G352.7-0.1 & Ia?& 7.5$\pm0.5$    & 4$\times$3     & 7.6$\pm$0.5     & 34(26-42)d$_{7.5}^2$         & 0.24(0.19-0.32)   & 61, 71, 72\\
           &    &                &                &                 & 0.11(0.09-0.14)d$_{7.5}^2$         & 3.2(2.4-4.6)   & -, -,72\\ 
G355.6-0.0 & ?  & 13$\pm2.6^{c}$ & 4$\times$3     & 13.2$\pm$2.7    & 4.6(2.3-10.0)d$_{13}^2$            & 0.56(0.55-0.57)   & -, 73, 73 \\
G359.1-0.5 & ?  & 5$\pm1^{c}$    & 12$\times$12   & 17.5$\pm$3.5    & 0.24(0.18-0.32)d$_{5}^2$           & 0.29(0.27-0.31)   & -, 74, 75\\
\enddata
\label{tab:TBLobserved}
\tablenotetext{a}{For SNRs with only one measured thermal plasma component, $EM$ and $kT$ are given; for SNRs with two measured thermal plasma components,
$EM_1$ and $kT_1$ are given in the first line and $EM_2$ and $kT_2$ are given in the second line.}
\tablenotetext{b}{The $EM$ and $kT$ quantities in parentheses are the 1$\sigma$ lower
and upper limits.}
\tablenotetext{c}{For SNRs without a quoted distance uncertainty, we adopt 20\% error.}
\tablenotetext{d}{In almost all cases, the radius error is dominated by the distance uncertainty.}
\tablenotetext{e}{15$\pm$5 kpc is adopted for G304.6+0.1; 13$\pm$7 kpc for  G311.5-0.3; 10$\pm$5 for G330.2+1.0.}
\tablenotetext{f}{References are for the SNe type, distances, and EM and kT values, respectively.}
\tablerefs {(1) \cite{2014Huang}, (2) \cite{2015Broersen}, (3) \cite{1998Giacani}, (4) \cite{2018Shan}, (5) \cite{1995Saken}, (6) \cite{2009Hui}, (7) \cite{2013Leahy} , (8) \cite{2004Mavromatakis}, (9) \cite{2012Leahy}, (10) \cite{2001Kothes}, (11) \cite{2008Jackson}, (12) \cite{2017Boubert}, (13) \cite{2006Lazendic}, (14) \cite{2017Nakano}, (15) \cite{2018SanchezCruces}, (16) \cite{2015Nakano}, (17) \cite{2004Sasaki}, (18) \cite{2004YarUyaniker}, (19) \cite{1994Hailey}, (20) \cite{2010Pannuti}, (21) \cite{2016Zhou}, (22) \cite{2009Katsuda}, (23)\cite{2016Katsuda}, (24) \cite{2007Leahy}, (25) \cite{1995Leahy}, (26) \cite{1989Landecker}, (27) \cite{2017Matsumura}, (28) \cite{2018Katsuda}, (29)  \cite{2017Reynoso}, (30) \cite{2013Katsuda}, (31) \cite{2012Sezer}, (32) \cite{2001Harrus}, (33) \cite{2013Prinz}, (34) \cite{1998Gaensler}, (35) \cite{2012SanchezAyaso}, (36) \cite{2007Park}, (37) \cite{2014Post}, (38) \cite{1975Caswell}, (39) \cite{2014Pannuti}, (40) \cite{2017Sezer}, (41) \cite{2019Sawada}, (42) \cite{2016combi}, (43) \cite{2012Prinz}, (44) \cite{2019Shan}, (45) \cite{2012Hui}, (46) \cite{2001Rakowski}, (47) \cite{2017Pannuti},  (48) \cite{2014Broersen}, (49) \cite{1996Rosado}, (50) \cite{2015Heinz}, (51) \cite{2013Heinz}, (52) \cite{2015Xing}, (53) \cite{2008Chen}, (54) \cite{1991ApJ...374..218L}, (55) \cite{1991ApJ...374..218L} , (56) \cite{2001McClureGriffiths}, (57) \cite{2009Park}, (58) \citep{2015Frank}, (59) \citep{2004Vink}, (60) \citep{2015Zhu}, (61) \citep{2014Yamaguchi}, (62) \citep{2006Rakowski}, (63) \citep{2015Zhang}, (64) \citep{1998Koralesky}, (65) \citep{2015Katsuda}, (66) \citep{2016Tsuji}, (67) \citep{2012Tian}, (68) \citep{2009Nakamura}, (69) \citep{2014Tian}, (70) \cite{2014Yasumi}, (71) \cite{2009Giacani}, (72) \cite{2014Pannuti1}, (73) \cite{2013Minami}, (74) \cite{2011Frail}, (75) \cite{2011Ohnishi}}
\end{deluxetable}

\clearpage

\begin{deluxetable}{crrrrrrrrrr}
\tabletypesize{\scriptsize}
\tablecaption{All SNRs: Standard (s=0, n=7 forward shock) Model Results} 
\tablewidth{0pt}
\tablehead{
\colhead{SNR}  & \colhead {$M_{ej}$} & \colhead{Age(+.-)} &  \colhead{Energy(+,-)}    & \colhead{Density(+,-)}   \\
\colhead{}   & \colhead{($M_{\odot}$)} & \colhead{(yr)}  & \colhead{($10^{50}$erg)}  & \colhead{($10^{-2}$ cm$^{-3}$)} & \\
}
\startdata
G38.7-1.3  	& 5	  &	8600(+6100,-2700)	&	1.27(+1.45,-0.84)	&	1.32(+0.43,-0.31)	   \\
G53.6-2.2   & 1.4 &	44600(+4600,-4600)	&	7.7(+2.9,-2.2)	&	9.6(+1.5,-1.2)	   \\
G67.7+1.8   & 5	  &	2300(+5300,-800)	&	0.28(+0.54,-0.07)	&	3.5(+1.7,-2.2)	   \\
G78.2+2.1   & 5	  &	9400(+2300,-1600)	&	6.3(+5.8,-3.7)	&	5.4(+2.1,-1.9)	   \\
G82.2+5.3  	& 1.4 &	17900(+3000,-2500)	&	8.3(+4.1,-2.5)	&	1.13(+0.42,-0.22)	   \\
G84.2-0.8  	& 1.4 &	10100(+1600,-1400)	&	1.82(+0.98,-0.79)	&	2.51(+0.44,-0.52)	   \\
G85.4+0.7   & 5	  &	5700(+2600,-2200)	&	1.79(+0.73,-0.59)	&	2.2(+1.4,-0.5)	   \\
G85.9-0.6   & 1.4 &	8000(+1900,-1600)	&	1.40(+1.42,-0.81)	&	1.36(+0.29,-0.17)	   \\
G89.0+4.7   & 5	  &	16400(+1600,-1300)	&	3.6(+1.8,-1.3)	&	0.71(+0.12,-0.09)	   \\
G109.1-1.0  & 5	  &	12700(+800,-800)    &	2.60(+0.60,-0.51)	&	35.6(+2.5,-2.1)     \\
G116.9+0.2  & 5   &	16700(+1400,-1300)	&	1.98(+0.55,-0.47)	&	17.5(+1.0,-0.9)     \\
G132.7+1.3  & 5 &	27900(+1300,-2700)	&	3.83(+0.85,-0.48)	&	10.9(+0.3,-0.3)     \\
G156.2+5.7  & 5	  &	29100(+7500,-7600)	&	61(+48,-32)&	20.4(+3.9,-2.6)     \\
G160.9+2.6  & 5	  &	5700(+3500,-3100)	&	2.2(+1.1,-1.1)	&	0.47(+0.23,-0.10)	   \\
G166.0+4.3  & 5 &	14500(+5600,-4100)	&	17(+17,-11)&	4.4(+4.2,-0.1)	   \\
G260.4-3.4  & 5	  &	3300(+1100,-1100)	&	1.34(+0.45,-0.41)	&	0.35(+0.10,-0.07)	   \\
G272.2-3.2  & 1.4 &	7500(+3800,-3300)	&	4.6(+7.6,-3.8)	&	14.1(+4.9,-2.7)     \\
G296.7-0.9  & 5 &	11600(+1300,-1300)	&	6.9(+3.9,-2.5)	&	18.0(+4.8,-3.3)     \\
G296.8-0.3  & 5	  &	10400(+600,-500)	          &	6.8(+1.6,-1.5)	&	2.23(+0.26,-0.27)	   \\
G299.2-2.9  & 1.4 &	8800(+2600,-2800)	&	0.74(+1.40,-0.45)	&	2.2(+1.6,-0.8)	   \\
G304.6+0.1  & 5 &	8900(+2300,-2100)	&	4.6(+5.3,-2.8)	&	4.2(+0.8,-0.6)	   \\
G306.3-0.9  & 1.4 &	12800(+2500,-2500)	&	10.2(+7.2,-4.8)	&	250(+59,-44)  \\
G308.4-1.4  & 5	  &	2900(+600,-500)	    &	0.44(+0.22,-0.15)	&	16.0(+2.3,-2.2)     \\
G309.2-0.6  & 5	  &	1100(+600,-600) 	&	3.4(+2.8,-2.5)	&	0.21(+0.23,-0.15)	   \\
G311.5-0.3  & 5 &	6900(+2900,-4400)	&	1.07(+2.01,-0.68)	&	6.2(+4.4,-1.3)	   \\
G315.4-2.3  & 1.4 &	11600(+800,-800)	&	8.5(+2.0,-1.8)	&	24.7(+2.8,-2.7)     \\
G322.1+0.0  & 5	  &	4300(+2800,-2000)	&	1.40(+3.65,-0.89)	&	5.0(+2.2,-1.4)	   \\
G327.4+0.4  & 5 &	10600(+1900,-1700)	&	3.0(+2.3,-1.6)	&	22.4(+4.8,-4.5)     \\
G330.0+15.0 & 5 &	37100(+15100,-13100)&	1.5(+2.8,-1.3)	&	4.1(+1.7,-1.3)	   \\
G330.2+1.0  & 5   &	9800(+5200,-6300)	&	2.0(+13.3,-1.2)	&	3.2(+3.7,-1.6)	   \\
G332.4-0.4 	& 5	  &	4000(+300,-800)	    &	0.83(+0.55,-0.22)	&	110.(+10.,-7.)   \\
G332.4+0.1  & 5 &	7000(+1600,-1400)	&	27(+18,-13)&	11.8(+3.0,-2.2)     \\
G332.5-5.6  & 5 &	10200(+1900,-1700)	&	0.67(+0.64,-0.37)	&	1.6(+1.1,-0.7)	   \\
G337.2-0.7  & 1.4 &	3100(+1300,-1900)	&	0.55(+1.55,-0.42)	&	44.(+27,-12)   \\
G337.8-0.1 	& 5	  &	3600(+1300,-1200)	&	10.6(+12.4,-4.8)	&	5.7(+2.6,-1.8)	   \\
G347.3-0.5  & 5	  &	6800(+1100,-2000)	&	0.73(+0.80,-0.22)	&	6.8(+9.4,-3.8)	   \\
G348.5+0.1  & 5   &	11400(+2200,-2100)	&	8.8(+7.0,-4.7)	&	21.6(+5.8,-5.6)     \\
G348.7+0.3 	& 5   &	14900(+1900,-1800)	&	29(+13,-9)	&	5.1(+1.4,-0.9)	   \\
G349.7+0.2  & 5   &	2900(+200,-200)	    &	1.50(+0.74,-0.55)	&	343(+55,-56)  \\
G350.1-0.3  & 5   &	2500(+500,-500)	    &	0.60(+0.34,-0.17)	&	428(+91,-75)  \\
G352.7-0.1  & 1.4 &	7600(+900,-1000)	&	2.44(+0.81,-0.64)	&	166(+25,-25)  \\
G355.6-0.0  & 5 &	9000(+1700,-1700)	&	4.9(+6.5,-3.0)	&	26(+17,-9)    \\
G359.1-0.5  & 5 &	18700(+2000,-1700)	&	0.84(+0.73,-0.41)	&	3.6(+0.3,-0.2)	   \\
\enddata
\label{tab:TBLstdmodel}
\end{deluxetable}

\clearpage

\begin{deluxetable}{crrrrrrrrrr} 
\tabletypesize{\scriptsize}
\tablecaption{Mixed Morphology SNRs: WL$^{a}$ Model Results} 
\tablewidth{0pt}
\tablehead{
\colhead{SNR} & \colhead{Age(+,-)} & \colhead{Energy(+,-)} & \colhead{$n_0$(+,-)} \\
\colhead{} & \colhead{(yr)} & \colhead{($10^{50}$erg)} & \colhead{($10^{-2}$cm$^{-3}$)} \\
}
\startdata
G38.7-1.3  & 5600(+4400, -1800)  & 1.22(+1.45, -0.84)  & 0.90(+0.29,	-0.21)	\\
G53.6-2.2  & 45900(+4800, -4800) & 7.6(+3.0, -2.3)  & 5.55(+0.82,	-0.67)	\\
G82.2+5.3  & 16600(+3000, -2500) & 9.4(+4.4, -2.8)  & 0.64(+0.24,	-0.12)	\\
G166.0+4.3 & 13600(+5800, -4600) & 19(+18, -11)& 2.50(+2.46,	-0.10)	\\
G304.6+0.1 & 7100(+2800, -2700)  & 5.6(+5.5, -3.4)  & 2.56(+0.75,	-0.48)	\\
G311.5-0.3 & 4700(+3300, -3000)  & 1.12(+2.41, -0.99)  & 4.2(+3.1,	-1.2)	\\
G337.8-0.1 & 2500(+900, -800)    & 14.8(+16.3, -6.9) & 3.7(+1.6,	-1.1)	\\
\enddata
\label{tab:customWL}
\tablenotetext{a}{Model type is WL2, the model of \citet{1991WL} with parameter $C/\tau$=2, to which we have added Coulomb electron-ion equilibration.}
\end{deluxetable}

\clearpage

\begin{deluxetable}{crrrrrrrrrrr} 
\tabletypesize{\scriptsize}
\rotate
\tablecaption{SNRs with Two Components: Predicted Reverse Shock Properties} 
\tablewidth{0pt}
\tablehead{
\colhead{SNR/Type$^{a}$} & s,n & \colhead{Age} & \colhead{Energy} & \colhead{$kT_{r}$} & \colhead{$EM_{r}$} 
& \colhead{$n_0$} & \colhead{$\rho_s$} \\ 
\colhead{} & \colhead{} & \colhead{(yr)} & \colhead{($10^{50}$erg)} & \colhead{(keV)} & \colhead {($10^{56}$cm$^{-3}$)} & \colhead{cm$^{-3}$} & \colhead{($10^{13}M_{\odot}$s/(km~yr))} \\
}
\startdata
G53.6-2.2/Ia & 0,7 & 44600(+4600,-4600) & 7.6(+2.9,-2.2) & 4.77(+1.01,-0.88) & 6.8(+1.9,-1.4) $\times 10^{-5}$  & 0.096(+0.015,-0.012) & n/a\\
 & 0,10 & 44200(+4600,-4600) & 7.8(+3.0,-2.3) & 2.97(+0.63,-0.55) & 11.7(+3.4,-2.4)$\times 10^{-5}$ & 0.096(+0.015,-0.012) & n/a\\
 & 2,7 & 6400(+900,-900) & 36.6(+5.4,-5.1) & 1.27(+0.15,-0.14) & 0.30(+0.13,-0.09) & n/a & 18.3(+4.7,-3.9)\\
 & 2,10 & 13200(+1800,-1800) & 3.2(+0.47,-0.43) & 0.152(+0.020,-0.018) & 3.9(+1.7,-1.2) & n/a & 23.0(+5.9,-4.9)\\
G109.1-1.0/CC & 0,7 & 12700(+800,-800) & 2.60(+0.59,-0.50) & 0.380(+0.077,-0.066) & 4.29(+0.73,-0.28) & 0.356(+0.025,-0.020) & n/a\\
 & 0,10 & 12600(+800,-800) & 2.64(+0.60,-0.51) & 0.190(+0.072,-0.028) & 6.14(+1.67,-0.164) & 0.357(+0.024,-0.020) & n/a\\
 & 2,7 & 1400(+100,-100) & 79.9(+6.3,-6.2) & 0.266(+0.010,-0.010) & 243(+54,-44) & n/a & 8.9(+1.3,-1.1)\\
 & 2,10 & 2800(+300,-200) & 14.5(+1.2,-1.15 & 0.053(+0.002,-0.002) & 3150(+700,-570) & n/a & 11.2(+1.6,-1.4)\\
G116.9+0.2/CC & 0,7 & 16700(+1400,-1300) & 1.98(+0.55,-0.46) & 0.292(+0.072,-0.062) & 2.25(+0.81,-0.25) & 0.175(+0.010,-0.009) & n/a\\
 & 0,10 & 16500(+1400,-1400) & 2.01(+0.55,-0.47) & 0.146(+0.051,-0.020) & 3.47(+1.34,-0.53) & 0.175(+0.009,-0.008) & n/a\\
 & 2,7 & 1800(+200,-200) & 70(+16,-15) & 0.229(+0.031,-0.031) & 106(+23,-21) & n/a & 6.51(+1.04,-0.98)\\
 & 2,10 & 3500(+600,-400) & 13.5(+4.0,-3.5) & 0.046(+0.006,-0.006) & 1370(+310,-270) & n/a & 8.1(+1.3,-1.2)\\
G156.2+5.7/CC & 0,7 & 29100(+7500,-7600) & 61(+48,-32) & 4.2(+1.7,-1.5) & 0.73(+0.60,-0.27) & 0.204(+0.039,-0.026) & n/a\\
 & 0,10 & 28800(+7500,-7400) & 62.6(+49.3,-33.2 & 2.614(+1.075,-0.940) & 1.25(+1.05,-0.47) & 0.204(+0.039,-0.026) & n/a\\
 & 2,7 & 3200(+900,-900) & 560(+160,-160) & 0.417(+0.031,-0.031) & 2200(+1300,-1000) & n/a & 46(+20,-17)\\
 & 2,10 & 6200(+1900,-1800) & 59.6(+9.4,-8.8) & 0.084(+0.006,-0.006) & 28000(+17000,-13000) & n/a & 58(+25,-21)\\
G299.2-2.9/Ia & 0,7 & 8800(+2600,-2800) & 0.74(+1.40,-0.45) & 1.26(+1.75,-0.40) & 17.3(+33.6,-9.9) $\times 10^{-5}$ & 0.022(+0.016,-0.007) & n/a\\
 & 0,10 & 8400(+2800,-2000) & 0.80(+1.38,-0.60) & 0.94(+0.61,-0.34) & 2.8(+6.0,-1.7) $\times 10^{-4}$ & 0.023(+0.016,-0.008) & n/a\\
 & 2,7 & 700(+300,-300) & 56(+40,-27) & 8.5(+1.9,-2.0) & 12.2(+36.6,-9.4) $\times 10^{-4}$ & n/a & 0.74(+0.98,-0.46)\\
 & 2,10 & 1100(+500,-400) & 22.4(+6.5,-6.5) & 1.65(+0.40,-0.40) & 0.016(+0.047,-0.012) & n/a & 0.93(+1.24,-0.58)\\
G306.3-0.9/Ia & 0,7 & 12800(+2500,-2500) & 10.2(+7.2,-4.8) & 6.4(+2.4,-2.0) & 18.0(+11.0,-5.9)$\times 10^{-4}$ & 2.50(+0.59,-0.44) & n/a\\
 & 0,10 & 12700(+2500,-2500) & 10.4(+7.4,-4.9) & 3.9(+1.5,-1.3) & 3.1(+1.9,-1.0) $\times 10^{-3}$& 2.51(+0.59,-0.42) & n/a\\
 & 2,7 & 2700(+700,-600) & 22.2(+4.6,-4.45) & 0.972(+0.078,-0.063) & 7.6(+5.6,-3.6) & n/a & 53(+23,-18)\\
 & 2,10 & 5000(+1100,-1100) & 2.55(+0.31,-0.32) & 0.131(+0.008,-0.005) & 99(+72,-46) & n/a & 67(+29,-23)\\
G315.4-2.3/Ia & 0,7 & 11600(+800,-800) & 8.4(+2.0,-1.8) & 8.6(+1.1,-1.1) & 35.1(+6.1,-5.1)$\times 10^{-5}$ & 0.248(+0.028,-0.027) & n/a\\
 & 0,10 & 11600(+800,-700) & 8.5(+2.1,-1.8) & 5.29(+0.73,-0.67) & 60.3(+10.5,-8.8)$\times 10^{-5}$ & 0.247(+0.028,-0.027) & n/a\\
 & 2,7 & 1100(+100,-100) & 161(+27,-27) & 6.98(+0.98,-0.98) & 0.225(+0.069,-0.058) & n/a & 11.1(+2.0,-1.8)\\
 & 2,10 & 1900(+200,-200) & 26.0(+5.6,-5.3) & 1.16(+0.19,-0.19) & 2.91(+0.89,-0.75) & n/a & 13.9(+2.5,-2.3)\\
G330.0+15.0/? & 0,7 & 35800(+15700,-15100) & 1.5(+2.6,-1.2) & 0.65(+0.51,-0.50) & 0.089(+0.116,-0.049) & 0.041(+0.020,-0.014) & n/a\\
 & 0,10 & 35600(+15500,-15100) & 1.5(+2.7,-1.2) & 0.40(+0.33,-0.32) & 0.154(+0.142,-0.086) & 0.041(+0.020,-0.014) & n/a\\
 & 2,7 & 4000(+2100,-2000) & 47(+23,-23) & 0.141(+0.002,-0.002) & 29(+51,-23) & n/a & 4.6(+4.5,-3.0)\\
 & 2,10 & 8000(+4600,-4100) & 8.6(+2.0,-2.1) & 0.028(+0.000,-0.000) & 390(+660,-310) & n/a & 5.8(+5.7,-3.8)\\
G332.4-0.4/CC? & 0,7 & 4000(+300,-800) & 0.83(+0.55,-0.23) & 0.311(+0.022,-0.021) & 63(+33,-20) & 1.10(+0.10,-0.07) & n/a\\
 & 0,10 & 3800(+300,-200) & 0.89(+0.59,-0.41) & 0.173(+0.039,-0.033) & 112(+51,-37) & 1.134(+0.099,-0.070) & n/a\\
 & 2,7 & 300(+100,-100) & 107(+25,-24) & 0.595(+0.042,-0.042) & 138(+67,-52) & n/a & 4.0(+1.4,-1.2)\\
 & 2,10 & 500(+100,-100) & 37.1(+4.6,-4.7) & 0.120(+0.008,-0.008) & 1780(+870,-680) & n/a & 5.1(+1.7,-1.5)\\
G349.7+0.2/CC & 0,7 & 2900(+200,-200) & 1.50(+0.73,-0.55) & 0.336(+0.040,-0.022) & 139(+45,-33) & 3.44(+0.54,-0.57) & n/a\\
 & 0,10 & 2700(+200,-200) & 1.61(+0.76,-0.59) & 0.226(+0.042,-0.038) & 234(+82,-58) & 3.55(+0.55,-0.57) & n/a\\
 & 2,7 & 200(+30,-30) & 118(+17,-17) & 0.626(+0.042,-0.042) & 680(+350,-270) & n/a & 8.1(+2.4,-2.2)\\
 & 2,10 & 500(+100,-100) & 31.7(+4.3,-4.1) & 0.126(+0.008,-0.008) & 8800(+4600,-3500) & n/a & 10.2(+3.0,-2.7)\\
G350.1-0.3/CC & 0,7 & 2500(+500,-500) & 0.60(+0.34,-0.17) & 0.258(+0.022,-0.022) & 290(+159,-92) & 4.28(+0.91,-0.75) & n/a\\
 & 0,10 & 2600(+300,-200) & 0.56(+0.45,-0.23) & 0.131(+0.033,-0.026) & 510(+220,-160) & 4.41(+0.85,-0.77) & n/a\\
 & 2,7 & 200(+50,-50) & 57(+13,-13) & 0.500(+0.042,-0.042) & 420(+320,-210) & n/a & 5.4(+2.4,-1.9)\\
 & 2,10 & 400(+100,-100) & 16.8(+3.0,-2.8) & 0.100(+0.008,-0.008) & 5450(+4100,-2700) & n/a & 6.7(+3.0,-2.4)\\
G352.7-0.1/Ia? & 0,7 & 7600(+900,-1000) & 2.44(+0.81,-0.64) & 3.08(+1.02,-0.64) & 31.4(+6.0,-5.2) $\times 10^{-4}$ & 1.65(+0.25,-0.25) & n/a\\
 & 0,10 & 7500(+900,-1000) & 2.48(+0.82,-0.64) & 1.89(+0.63,-0.39) & 54.3(+10.5,-9.0) $\times 10^{-4}$& 1.67(+0.25,-0.25) & n/a\\
 & 2,7 & 1100(+300,-300) & 23.6(+18.3,-9.0) & 2.01(+1.19,-0.67) & 0.90(+0.36,-0.30) & n/a & 14.8(+3.3,-3.1)\\
 & 2,10 & 2300(+600,-600) & 3.1(+2.3,-1.1) & 0.240(+0.157,-0.077) & 11.7(+4.7,-3.9) & n/a & 18.6(+4.1,-3.9)\\
\enddata
\label{tab:customS2}
\tablenotetext{a}{For CC-type or unknown-type, we use $M_{ej}=5M_{\odot}$ and CC-type ejecta abundances; for Ia or Ia?-type, we use$M_{ej}=1.4M_{\odot}$ and Ia-type ejecta abundances.}
\end{deluxetable}

\clearpage


\end{document}